\documentclass[aps,pre,twocolumn,groupedaddress,amsmath,amssymb,nofootinbib]{revtex4-1}

\usepackage{graphicx}
\usepackage{caption}
\usepackage{subcaption}
\usepackage{amsmath}
\usepackage{epstopdf}
\usepackage{soul}
\usepackage{upgreek}
\usepackage{calrsfs}
\usepackage[usenames, dvipsnames]{color} 
\usepackage{tipa}
\usepackage{upgreek}

\usepackage{tabularx}
\usepackage{tikz}
\usetikzlibrary{shapes, arrows}
\tikzstyle{line}=[draw] 
\tikzstyle{startstop} = [rectangle, rounded corners, minimum width=3cm, minimum height=1cm,text centered, draw=black, text width=2cm, fill=red!30]
\tikzstyle{io} = [trapezium, trapezium left angle=70, trapezium right angle=110, minimum width=3cm, minimum height=1cm, text centered, draw=black, fill=blue!30]
\tikzstyle{decision} = [diamond, minimum width=3cm, minimum height=1cm, text centered, draw=black, fill=green!30]
\tikzstyle{arrow} = [thick,->,>=stealth]
\tikzstyle{process} = [rectangle, minimum width=3cm, minimum height=1cm, text centered, text width=2cm, draw=black, fill=orange!30]

\newcommand{\be}{\begin{equation}}
\newcommand{\ee}{\end{equation}}
\newcommand{\bea}{\begin{eqnarray}}
\newcommand{\eea}{\end{eqnarray}}
\newcommand{\ba} {\begin{align} }
\newcommand{\ea} {\end{align} }

\newcommand{\kbt}{k_{\mathrm{B}}T}

\newcommand{\lb}{l_{{\rm B}}}
\newcommand{\ld}{\lambda_{{\rm D}}}

\newcommand{\mCpA}{{\rm mC/m}^2}

\newcommand{\rbf}{\mathbf{r}}
\newcommand{\qbf}{\mathbf{q}}
\newcommand{\pbf}{\mathbf{p}}

\begin{document}


\title{Breakdown of electroneutrality in nanopores}

\author{Amir Levy[1], J. Pedro de Souza[2], and Martin Z. Bazant[2,3]}
\affiliation{[1] Department of Physics, Massachusetts Institute of Technology, Cambridge, Massachusetts 02139 USA}
\affiliation{[2] Department of Chemical Engineering, Massachusetts Institute of Technology, Cambridge, Massachusetts 02139 USA}
\affiliation{[3] Department of Mathematics, Massachusetts Institute of Technology, Cambridge, Massachusetts 02139 USA}

\begin{abstract}
Ion transport in extremely narrow nanochannels has gained increasing interest in recent years due to unique physical properties at the nanoscale and the technological advances that allow us to study them. It is tempting to approach this confined regime with the theoretical tools and knowledge developed for membranes and microfluidic devices, and naively apply continuum models, such as the Poisson-Nernst-Planck and Navier-Stokes equations. However, it turns out that some of the most basic principles we take for granted in larger systems, such as the complete screening of surface charge by counter-ions, can break down under extreme confinement. We show that in a truly one-dimensional system of ions interacting with three-dimensional electrostatic interactions, the screening length is exponentially large, and can easily exceed the macroscopic length of a nanotube. Without screening, electroneutrality breaks down within the nanotube, with fundamental consequences for ion transport and electrokinetic phenomena. In this work, we build a general theoretical framework for electroneutrality breakdown in nanopores, focusing on the most interesting case of a one-dimensional nanotube, and show how it provides an elegant interpretation for the peculiar scaling observed in experimental measurements of ionic conductance in carbon nanotubes.

\end{abstract}

\maketitle

\maketitle

\section{Introduction}
The transport of ions in extreme confinement has applications ranging from physiology to chemical engineering\cite{hille1978ionic,doyle1998structure,berneche2001energetics,hess1984mechanism,werner2001streaming,vilatela2015tough,yang2003electrokinetic,kim2010direct,kim2010power,mani2011deionization,PhysRevE.93.053108,schmuck2015homogenization}. Whether we consider ions traveling through the protein channels in the cell membrane or through pores in an ion-exchange membrane, the underlying physics shares many similarities\cite{nonner1998ion,tedesco2016nernst}. A growing interest in ionic transport through nanopores has emerged in recent years owing to nano-fabrication advances that enable us to study pristine nano-channels at the single-channel level, such as carbon nanotubes, boron-nitrite nanotubes or silicon nano-channels\cite{lee2010coherence,agrawal2017observation,siria2013giant,guan2012electric,daiguji2004electrochemomechanical}. These experiments have revealed that our understanding of even the basic physics is incomplete, and important theoretical knowledge gaps still exist\cite{CENT}. 

Classical theories of ion transport were inspired by the membrane technology that was available at that time\cite{staverman1952non,gross1968membrane,fair1971reverse}. The complicated interplay of fluxes and potential gradients (chemical, electrical, and pressure) was naturally modeled with continuum theories. The so-called ``capillary-pore model," based on the Poisson-Boltzmann (PB) equation for the charge distribution  normal to the pore walls under the local equilibrium assumption, with Navier-Stokes and Nernst-Planck equations for the fluid flow and ionic flux, is a continuum linear response theory of transport in charged cylinders\cite{PhysRevE.93.053108}, which is widely used in different electrochemical applications, from electro-osmotic pumps to energy conversion devices\cite{sasidhar1981electrolyte,yao2003porous,petsev2006electrostatic,kim2010nanofluidic,chang2011electrokinetic,yan2013energy}. A competing school of thought, rooted in transition state theory\cite{zwolinski1949diffusion}, emerged in the biophysical community. Experiments on ion channels show that when open, the transport of the ions is best described by a discrete, single-file reaction model, where ions are attached to specific binding in the channel by a chemical reaction\cite{hille1978ionic,hess1984mechanism,almers1984non,doyle1998structure,berneche2001energetics,sather2003permeation}. 

Evidently, a new theory is required for nanochannels with a pore diameter of less than $10$nm (single-digit nanopores, or SDNs\cite{CENT}),  in order to span between the two limiting regimes, from discrete to continuum behavior. While the two traditional pictures have some merit, neither can exactly capture experimentally measured conductivity curves. Unusual scaling behavior of ionic conductance in carbon nanotubes (CNTs), for example, was recently reported by Secchi et al\cite{Secchi2016} and was subsequently interpreted with a Space-Charge continuum model\cite{Biesheuvel2016}. The conductance of a narrow CNT porin, in contrast, was fitted to a Michalis-Menten reaction model\cite{Tunuguntla2017}, suitable for a single-file transport mechanism. 

In this work, we propose a new theoretical framework for electrolytes in nanopores, consisting of confined ions with three-dimensional (3d) electric fields that extend into the surrounding matrix. In the most interesting and relevant case of SDNs, we construct a truly one-dimensional (1d) mean-field theory of ions confined to a long, thin nanopore in a 3d matrix, in contrast to previous models of ion chains with 1d Coulomb interactions [33-35]. Surprisingly, $1$d electrolytes exhibit several interesting behaviors, most notable is the breakdown of global electroneutrality:  a system can have a net charge where the total charge of the ionic solution does not exactly cancel out external charges. When the pore diameter is comparable to the spacing between ions, the system essentially behaves like a $1$d correlated electrolyte. Ions are not necessarily restricted to transport in single-file, but the nature of the electrostatic interactions resembles a $1$d chain.  Despite our interest in transport, this paper will only focus on the equilibrium properties of ion channels. The equilibrium properties can in turn be used to understand transport properties of the nanotube\cite{Biesheuvel2016, gross1968membrane}. 


Electroneutrality breakdown in nanopores has been observed in Monte-Carlo simulations \cite{sorensen1992ion,rivera1994grand,lo1995non,lee1997non,lee1999deviation,boda2000monte,boda2002monte,boda2006effect}, and was recently even measured experimentally\cite{luo2015electroneutrality}. However, it was not interpreted as a unique feature of the $1$d geometry. Instead, the breakdown was assumed to occur due to an excess screening of charges outside of the pore. This type of local breakdown of charge neutrality is not suited to most transport problems, where the channel is surrounded by a constant dielectric medium. 

Without charge neutrality, electric fields leak out of the confined region into the outer substrate. We derive (section II) a mean-field theory of a confined electrolyte by properly accounting for the outer region as well, and present and illustrate using numerical simulations the emergence of electroneutrality breakdown. In Section III we consider a uniformly charged pore and solve a self-consistent algebraic mean-field equation for the excess charge. Three length-scales govern the accumulated charge: the Debye screening length (ion-ion interactions), the Gouy-Chapman length (ion-wall interactions) and the pore diameter. In Section IV we present a general scaling argument for the enhanced screening length in low dimensions. We show that electroneutrality breakdown is a unique feature of $1$d systems due to their exponentially long screening length that can easily exceed the size of the system.  Finally (section V), we solve a full $1$d lattice mean-field equation and observe the emergence of ion-ion correlations at high concentrations. The breakdown of electroneutrality has profound implications on the transport of ions through nanochannels, and in section VI we show that our model can account for the unusual scaling of conductance in CNT.

\section{Mean-field theory of confined electrolytes}
Electroneutrality is often a hidden assumption of continuum models: it hides in the boundary conditions for the Poisson-Boltzmann equation,  where the electric fields outside the electrolyte are assumed to vanish \cite{Biesheuvel2016,PhysRevE.93.053108,gross1968membrane,fair1971reverse,sasidhar1981electrolyte,Secchi2016, yao2019strong, ren2008slip,cervera2005poisson, van2006electrokinetic,vlassiouk2008ionic,stein2004surface,li2015direct}. According to Gauss's law, if there is no electric flux emanating from the electrolyte in the pore, it has zero net charge.  Electroneutrality  relies on the nanometer-scale screening length, which guarantees macroscopic charge fluctuations are negligible. A rigorous analysis requires us to solve the Laplace equation outside the electrolyte, in addition to the PB equation inside. 

It is important to note that electroneutrality is not always assumed\cite{lozada1996violation,lozada1996violationb,luo2015electroneutrality,colla2016charge,schmuck2015homogenization}. As a recent example, Colla et al\cite{colla2016charge} considered two charged plates immersed in water, with free ions on both sides of each plate, and solved a density functional theory (DFT) in the entire space. Since the screening is not symmetric, especially when the screening length is large, the accumulated charge between the plates can be small. The authors consider this as an example of a local breakdown of electro-neutrality (LEC), and while it shares some similarities with our approach, LEC is fundamentally different from the pore-wide electroneutrality breakdown which we discuss, and is not a unique property of a $1$d geometry.  

	\subsection{ General equations }
\begin{figure}[htp]
  \includegraphics[width=\columnwidth ]{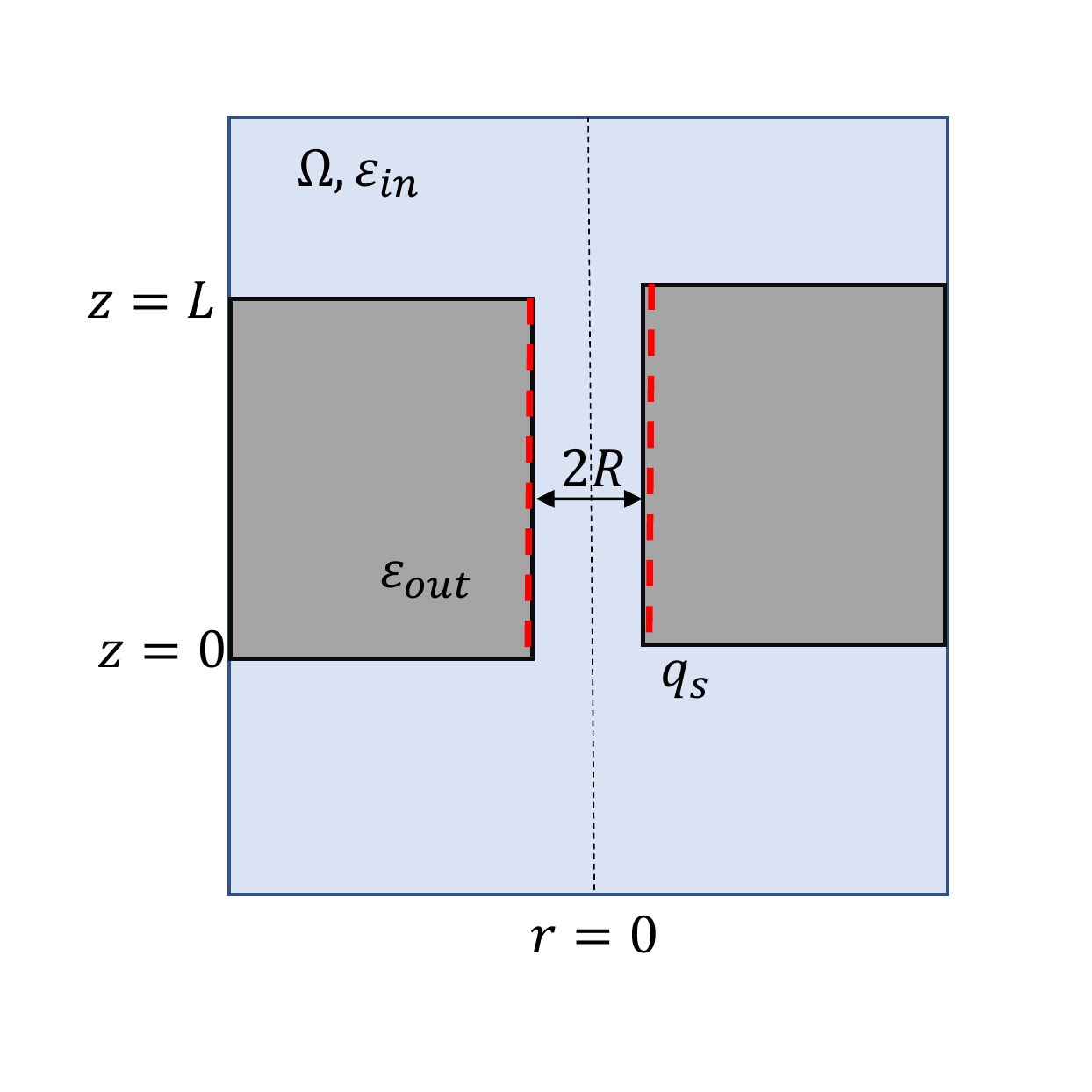}
  \caption{Sketch of a confined electrolyte,  in chemical equilibrium with a distant external reservoir (in the grand canonical ensemble), as indicated by an ideal ``salt bridge," which does not otherwise perturb the system. The governing equation in the inner region ($\Omega$) is the usual Poisson-Boltzmann equation. The outer region has a fixed dielectric constant and is described by a Laplace equation. The surface charge on the boundary layer ($\partial \Omega$) determines the jump in the normal component of the electric field.}
  \label{fig_sketch}
\end{figure}

Let us consider the PB equation for a symmetric binary monovalent electrolyte, fixed at a chemical potential that is set by an external reservoir with ionic concentration $c_0$, and confined to a small region in space ($\Omega$, see Fig.~\ref{fig_sketch}). We further assume that the electrolyte is embedded in a constant dielectric medium with permittivity $\varepsilon_{\rm out}$, and the boundary is charged with a surface charge $q_s$. The electrostatic potential ($\phi$) is determined by a set of PB and Laplace equations:
\begin{eqnarray}
\label{PBL_eqns}
\begin{cases}
    \varepsilon_{\rm in}\nabla^2 \phi_{\rm in}(\rbf) =  2 c_0 e\sinh\left[e\beta \phi_{\rm in}(\rbf)\right] & \forall\rbf \in \Omega \\
    \varepsilon_{\rm out}\nabla^2 \phi_{\rm out} (\rbf) = 0 & \forall\rbf \not\in \Omega.
\end{cases}
\end{eqnarray}
where $e$ is the electron charge, $\varepsilon_{\rm in}$ and $\varepsilon_{\rm out}$ are the dielectric constants in the solvent and dielectric matrix, respectively, and $\beta=1/\kbt$ is the inverse temperature and $k_{\rm B}$ is the Boltzmann constant. A similar approach was previously introduced to calculate the effect of image charges on the ionic self energy in confinement\cite{jordan1989electrolyte}, and to study the transport of ions through porous media\cite{schmuck2015homogenization}, but the resulting electroneutrality breakdown was not emphasized. 

For a cylindrical geometry, with charge density $q_s$ on the pore walls, the boundary conditions for this system are:
\begin{eqnarray}
\label{PBL_BC}
\left[\phi_{\rm out}(\rbf)-\phi_{\rm in}(\rbf)\right]_{\forall \rbf \in \partial\Omega} &=& 0  \nonumber\\ 
{\bf n}(\rbf) \cdot\left[\varepsilon_{\rm in}\nabla \phi_{\rm in}-\varepsilon_{\rm out}\nabla \phi_{\rm out}(\rbf)\right]_{\forall 0<r<R, 0<z<L}  &=&  q_s \nonumber\\
\left.\phi_{\rm out}(\rbf)\right|_{r\rightarrow \infty }&=&0,
\end{eqnarray}
where $\partial \Omega$ is the electrolyte boundary, and ${\bf n}$ is an outward unit vector normal to the boundary. 

 Solving the set of PB/Laplace equations for a finite cylinder can only be done numerically. In many cases, we will find that accounting for outer electric fields is redundant, since the outer electric fields vanish. In other cases, dramatic differences in the charge profile can be observed. In the the remaining sections of the paper, we will present simplified models that permit analytical results, which will help us explore the implication of Eq.~\ref{PBL_eqns}. While some of the accuracy of the complete model is lost, we will gain a much better physical understanding, as well as mathematically convenient approximations. 
\begin{figure}[htp]
  \includegraphics[width=\columnwidth ]{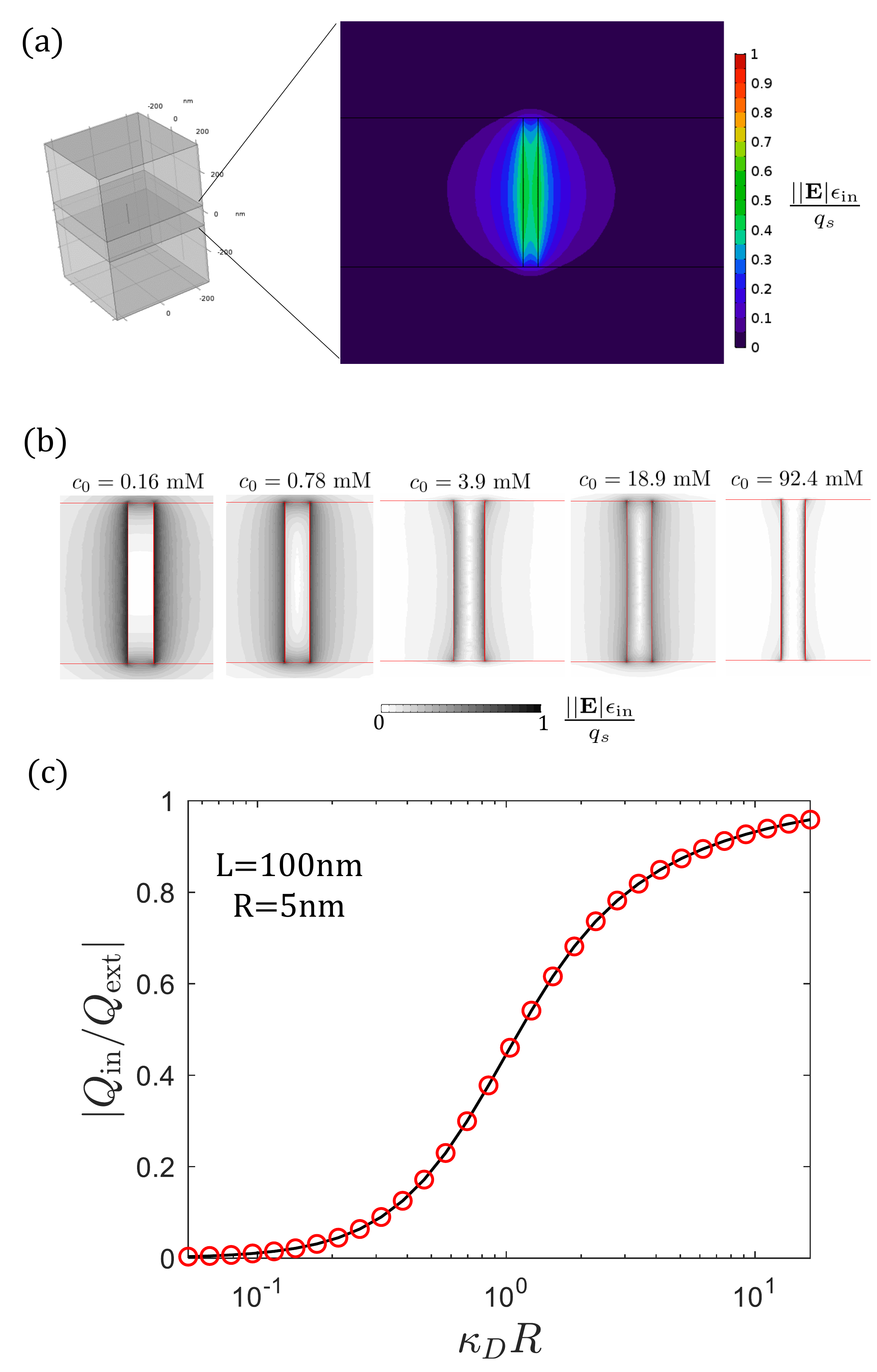}
  \caption{Electroneutrality breakdown for a nanopore in equilibrium with an external reservoir.  The surface charge is not fully screened by the confined electrolyte, as electric fields extend into the solid matrix (assumed here to have the same dielectric constant as the pore). (a) simulation box in COMSOL, as well as field intensity. (b) an illustration of the electric field intensity for different concentrations. As the concentration increases, the electric fields are screened from the outer region and are concentrated in a narrow region near the surface of the sphere. (c) the accumulated charge inside the sphere, relative to the total surface charge, as a function of the ratio of the radius to the Debye length }
  \label{fig_comsol}
\end{figure}

\subsection{Numerical simulation}

Remarkably, despite numerous theoretical investigations of ionic channels, there is currently no analysis of this simple and fundamental problem, to the best of our knowledge. In many cases, the full set of PB/Laplace equations is oversimplified by disregarding the outer dielectric matrix\cite{Biesheuvel2016,PhysRevE.93.053108,gross1968membrane,fair1971reverse,sasidhar1981electrolyte}. 

We solve equation \ref{PBL_eqns} using the COMSOL Multiphysics simulation package.  Fig.~\ref{fig_comsol}  shows an example of such numerical simulation for a nanopore configuration and the resulting electrostatic potential in a cross section of the cylinder. The simulation box is composed of a membrane dielectric embedding a charged pore that is between two reservoirs. The reservoirs have a thickness of $20\lambda_D$ and each side of the box is $5L$. At the top and the bottom of the box, Dirichlet boundary conditions are imposed, to signify a return to bulk reservoir conditions with $\phi=0$ and $c_i=c_0$. At the lateral surfaces, symmetry conditions are imposed. For simplicity, we analyze the limit of small surface charge density, $q_s$, on the pore walls. We also consider the case of $\varepsilon_{in}=\varepsilon_{out}=80\varepsilon_0$. Note that there is a significant `leakage' of the electric field into the dielectric matrix, demonstrating the invalidity of the local electroneutrality assumption, as illustrated in Fig. 2b. In Fig. 2c, we quantify the extent of electroneutrality breakdown by calculating the ratio of ionic charges within the cylinder, $Q_{in}$, to the number of charges on the pore walls, $Q_{ext}$. For $\mid Q_{in}/Q_{ext}\mid\rightarrow 1$, the pore walls and the pore are electroneutral, while for  $\mid Q_{in}/Q_{ext}\mid\rightarrow 0$, the pore walls and the pore domain experience electroneutrality breakdown. 

Note that the continuum simulation as a whole (pore walls, pore, and reservoirs) satisfies global electroneutrality. In fact, one way to check that the reservoir domains are sufficiently large is to ensure that the overall system satisfies the global electroneutrality constraint.  When electroneutrality breakdown occurs within the pore domain, the excess screening charge exists at the membrane-reservoir interfaces. Even so, the breakdown in electroneutrality within the pore domain remains an unexpected phenomenon, especially as the pore length goes to macroscopic scales many times longer than the bulk Debye length. Through a series of analytical approximations, we will show that electroneutrality breakdown arises especially strongly in 1d confinement, where the screening length becomes exponentially long.

\section{Uniform Embedded Pore model}

 The numerical solution of Eq.~\ref{PBL_eqns} in a long and narrow cylinder showed that overlapping double layers are accompanied by a net charge of the pore. With this result in mind, let us now introduce an analytical model of a uniformly charged pore embedded in a constant dielectric medium. This ``Uniform Embedded Pore" (UEP) model is closely related to the widely used ``Uniform Potential" (UP) model\cite{schlogl1955theorie,sonin1976osmosis,cwirko1989transport,verbrugge1990ion,bowen2002modelling,kim2010power,PhysRevE.93.053108}, also known as Teorell-Meyers-Sievers theory \cite{cowan1937general,teorell1953transport,meyer1936permeabilite}, which takes advantage of the narrow pore geometry to approximate a constant charge distribution in the radial direction. This approximation is further adapted in the ``Leaky membrane model", which describes the concentration polarization in porous media, based on similar microscopic assumptions\cite{mani2011deionization,yaroshchuk2012makes,dydek2013nonlinear}. While we make the same assumption about the charge distribution within the pore, we recognize that the sourrounding dielectric matrix cannot be neglected. At first, to keep matters simple, we only consider the case where the dielectric constants inside and outside of the pore are equal. 

\subsection{Derivation}
The free energy functional of $M$ ionic species immersed in a dielectric continuum with permittivity $\varepsilon$, assuming ideal mixing entropy, with  thermal de Broglie wavelength $\lambda_T$, reads:
\begin{eqnarray}
F[\{c_i(\rbf)\}]&=&\frac{1}{2}\int_V{\rm d}^3\rbf\int_V{\rm d}^3\rbf' \frac{\rho(\rbf)\rho(\rbf')}{4\pi\varepsilon|\rbf-\rbf'|} 
\nonumber\\
&+&\kbt \int_V  {\rm d}^3\rbf\sum_{i=1}^M c_i(\rbf) \left[\log \left(\lambda_{T}^3c_i(\rbf)\right)-1\right]
\nonumber\\
\rho(\rbf)&=&\sum_{i=1}^M e z_i c_i(\rbf) +\rho_{\rm ext},
\end{eqnarray}
where $c_i(\rbf)$ and $z_i$ are the concentration and valency of the $i$-th ionic species, respectively, and $\rho_{\rm ext}$ is an external charge distribution. For a uniform ionic density we write the free energy as a function of the mean ionic concentrations:
\begin{eqnarray}
F[\{c_i\}]&=&  \frac{e^2}{2}\sum_{i=1}^M\sum_{j=1}^M z_i z_j c_ic_j\int_V{\rm d}^3\rbf \int_V{\rm d}^3\rbf'\frac{1}{4\pi\varepsilon|\rbf-\rbf'|} \nonumber\\
&+& \int_V{\rm d}^3\rbf \int_V{\rm d}^3\rbf'\frac{\rho_{\rm ext}(\rbf) \rho_{\rm ext}(\rbf') }{4\pi \varepsilon|\rbf-\rbf'|}
\nonumber\\
&+& \sum_{i=1}^M e z_i c_i\int_V{\rm d}^3\rbf \int_V{\rm d}^3\rbf'\frac{\rho_{\rm ext}(\rbf)}{8\pi\varepsilon|\rbf-\rbf'|}
\nonumber\\
& + &  V\kbt\sum_{i=1}^M c_i \left[\log \left(\lambda_T^3 c_i\right)-1\right]
\end{eqnarray}
The ``mean-interaction" integrals in the above equation describe the electrostatic energy associated with uniformly distributed charge inside a volume, $V$. They depend only on the external charge and geometry, so it is convenient to define the following integrals:
\begin{eqnarray}
\label{MeanInteractionInt}
    \gamma &=& \frac{\lb}{e V}\int_V{\rm d}^3\rbf \int_V{\rm d}^3\rbf'\frac{1}{|\rbf-\rbf'|},
    \nonumber\\
    \rho_{\rm ext}^0 &=& \frac{\lb}{e\gamma V}\int_V{\rm d}^3\rbf \int_V{\rm d}^3\rbf'\frac{ \rho_{\rm ext}(\rbf)}{|\rbf-\rbf'|}.
\end{eqnarray}
The free energy density is now simplified:
\begin{equation}
\begin{aligned}
\frac{\beta F[\{c_i\}]}{V}&= \sum_{i=1}^M \Bigg[ \gamma z_i c_i   \left(\sum_{j=1 }^M\frac{ez_jc_j}{2}  +  \rho_{\rm ext}^0\right) \\
 &+  c_i \left(\log (\lambda_T^3 c_i)-1\right)\Bigg],
\end{aligned}
\end{equation}
where we neglect constant contributions to the free energy. The chemical potential is the derivative of the free energy density with respect to concentration:
\begin{eqnarray}
\label{muDef1}
\beta \mu_i &=& \gamma z_i \left(\sum_{j=1}^M  e z_j c_j + \rho_{\rm ext}^0\right)  +\log \left(\lambda_T^3 c_i\right).
\end{eqnarray}
 The first term in the LHS of Eq.~\ref{muDef1} describes the excess free energy due to electrostatic interactions, while the second term is the ideal gas entropy. If our system is in chemical equilibrium with a bulk reservoir, the chemical potential will only have the second term ($\beta \mu_i^0= \log \left(\lambda_T^3 c_i^0\right)$), with the bulk values of ionic concentrations. Equating the chemical potential in Eq.~\ref{muDef1} to the bulk value, we find the following equation for the average ionic density, $c_i$:
\begin{eqnarray}
\label{rhoEqn1}
\log \left(\lambda_T^3 c_i^0\right) &=& \gamma z_i \left(\sum_{j=1}^M  e z_j c_j + \rho_{\rm ext}^0\right)  +\log \left(\lambda_T^3 c_i\right)
\nonumber\\
c_i &=& c_i^0 \exp\left[\gamma z_i \left(\sum_{j=1}^M  e z_j c_j + \rho_{\rm ext}^0\right)\right]. 
\end{eqnarray}

The average charge density, $\rho^0 = \sum e z_i c_i$, therefore, satisfies the following equation:
\begin{eqnarray}
\rho^0&=&\sum_{i=1}^M e z_i c_i^0 \exp\left[-z_i \gamma (\rho_{\rm ext}^0+\rho^0)\right],
\end{eqnarray}
We can incorporate non-idealities to the system by adding an excess chemical potential to Eq.~\ref{muDef1}. If this were the case, the $c_i^0$ would be replaced by the ionic activities, rather than the concentrations. 

The equation for the mean charge in the UEP model, Eq.~\ref{rhoEqn1}, is written in terms of the average charge density, $\rho^0$, which makes it easier to solve. To see its relation with the set of PB/Laplace equations discussed in section II, we write Eq.~\ref{rhoEqn1} in terms of the electrostatic potential. Note that the term in the exponent is the electrostatic potential, averaged over the volume of the electrolyte, and multiplied by $z_i e\beta$. Furthermore, the Poisson equation relates charge density and electrostatic potential, $ \rho^0 = -\varepsilon\nabla^2 \phi$, and thus Eq.~\ref{rhoEqn1} becomes a partial differential equation:
\begin{eqnarray}
\label{PB_UEP}
\varepsilon \nabla^2 \phi(\rbf)=-\sum_{i=1}^M e z_i c_i^0 \exp\left[-z_i e\beta \overline{\phi(\rbf)}\right].
\end{eqnarray}
Hence, the uniform embedded pore model is an approximation of the standard PB equation, where we replace the electrostatic potential with its volume average. The complete set of equations for the potential in the UEP model includes in addition the Laplace equation in the outer dielectric medium, and the boundary conditions of Eq.~\ref{PBL_BC}. This set of equations is mathematically equivalent to Eq.\ref{rhoEqn1}, as long as the permittivity is everywhere the same.  

Under electroneutrality the charge density, $\rho^0$, exactly cancels out the external charge distribution, $\rho_{\rm ext}^0$. A criterion for electroneutrality to be satisfied, for weakly charged systems, is obtained by linearzing equation \ref{rhoEqn1}:
\begin{eqnarray}
\label{rhoEqn1Lin}
\rho^0 = -\frac{\rho_{\rm ext}^0}{1+(e \gamma \sum_{i=1}^M z_i^2 c_i^0)^{-1}}.
\end{eqnarray}
Hence, the condition for electroneutrality is $e\gamma \sum_{i=1}^M z_i^2 c_i^0 \gg 1$. Using the definition of $\gamma$ (Eq.~\ref{MeanInteractionInt}), we can write this condition in terms of the bulk Debye length- $\lambda_D^{-2} =4\pi \lb (\sum_{i=1}^M z_i^2 c_i^0)$:
\begin{eqnarray}
\label{ENcond}
\int_V \frac{{\rm d}^3\rbf}{V} \int_V \frac{ {\rm d}^3\rbf'}{\lambda_D^2 |\rbf-\rbf'|}\gg 1
\end{eqnarray}
In typical $2$ and $3$ dimensional systems, this condition is satisfied if the characteristic size of the system is larger than the Debye length,  but in $1$d we find a very different result.   
\begin{figure*}[!]
  \includegraphics[width=\linewidth ]{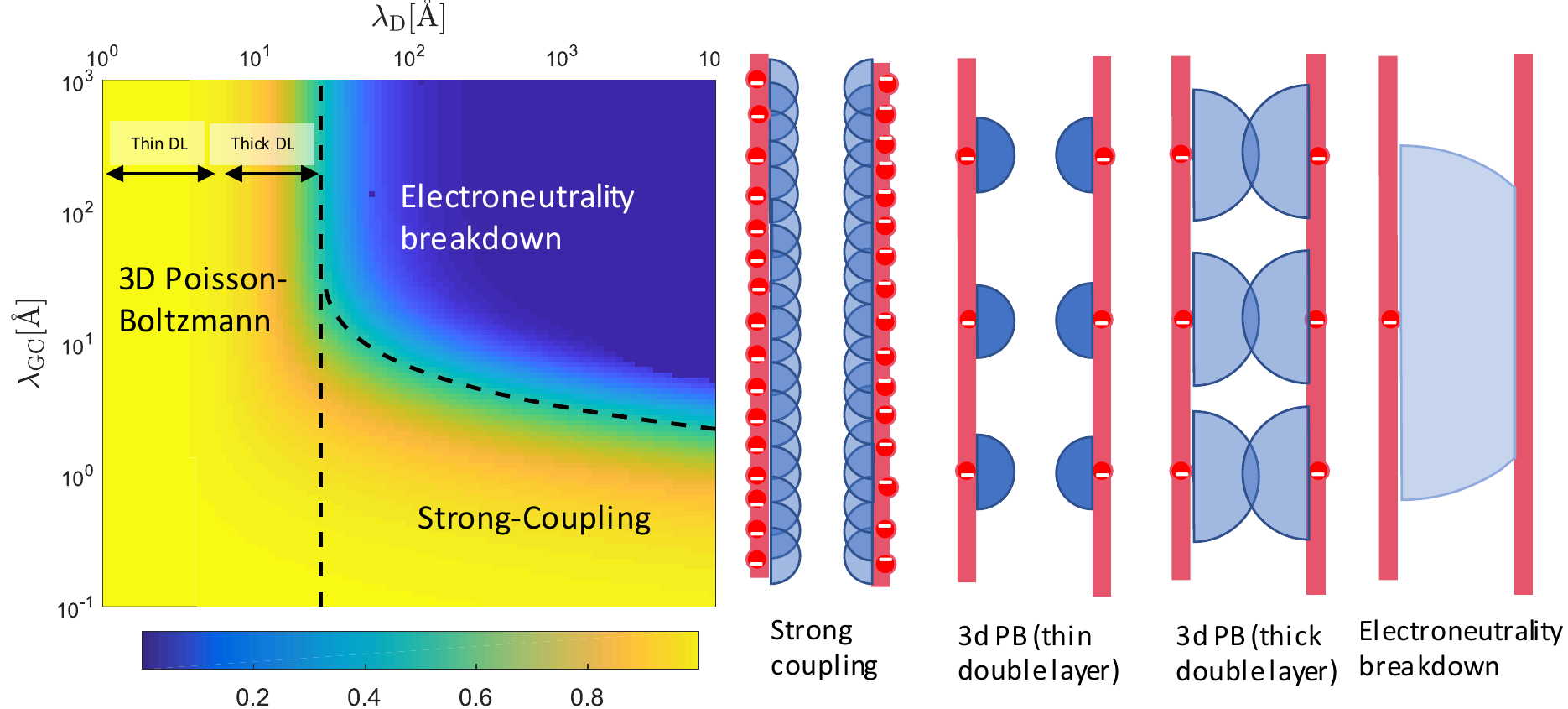}
  \caption{Left: Phase diagram for electroneutrality, as a function of the Gouy-Chapman length ($\lambda_{\rm GC}$) which is related to the surface charge, and Debye screening length ($\lambda_D$), which is related to the bulk ionic concentration. The color intensity indicates the ratio of total accumulated charge inside a nanochannel to the surface charge and is obtained by solving Eq. \ref{EqRhoSinh}. In electroneutral systems, this ratio is $1$. The dashed mark the transition to electroneutrality Eq.~(\ref{vPhaseDiagramBoundary}-\ref{phaseDiagramBoundary}). Right: illustration of the 4 regimes of the phase diagram. The blue circles mark the Debye screening length.}
  \label{fig_UEPpd}
\end{figure*}
\subsection{Cylindrical nanopore geometry: a phase diagram}
Our numerical solution suggests that electroneutrality breaks down for cylindrical nanopores if the Debye length is greater than the pore radius. We can now see that this is a property of the cylindrical mean-interaction integral. Let us consider a cylinder with radius $R$, length $L$, and a surface charge density $q_s$ ($\rho_{\rm ext} = q_s \delta(r-R)$). The mean-interaction integrals, $\gamma$, and $\rho_{\rm ext}$, can be approximated to a good precision by considering a test charge located at the center of the pore:
\begin{eqnarray}
\label{gammaDef}
 \gamma^{\rm apprx} &=& \frac{\lb}{e V}\int_V{\rm d}^3\rbf \int_V{\rm d}^3\rbf'\frac{1}{|\rbf|}
 \nonumber\\ 
 &=& \frac{\lb}{e}\int_0^R\int_{-L/2}^{L/2}\frac{2\pi r \, dr dz}{\sqrt{r^2+z^2}}=\frac{2\pi R^2 \lb}{e} \log\left(\frac{L}{2R}\right),
 \nonumber\\
 \rho_{\rm ext}^{0,{\rm apprx}} &=& \frac{\lb}{e\gamma^{\rm apprx} V}\int_V\frac{{\rm d}^3\rbf}{|\rbf|} \int_V{\rm d}^3\rbf'q_s \delta(r'-R).
 \nonumber\\
 &=&\frac{2\pi R L q_s}{V} = \frac{2 q_s}{R}.
\end{eqnarray}
Interestingly, this approximation deviates by only a few percents from an accurate numerical evaluation of the mean-interaction integrals. Based on this approximation, the electroneutrality condition ($e\gamma^{\rm apprx} \sum z_i^2 c_i^0 \gg 1$) in nanopores can be expressed in terms of the natural system length scales:
\begin{eqnarray}
\label{ENcondition}
 L \gg (2R)\exp\left(\frac{2\lambda_D^2}{R^2}\right),
\end{eqnarray}
which will be derived via scaling arguments in section IV. To extend the electroneutrality condition beyond weakly charged systems, the full solution of Eq.~\ref{rhoEqn1} is required. For a monovalent binary electrolyte, Eq.~\ref{rhoEqn1}   reads:
\begin{eqnarray}
\label{EqRhoSinh_0}
\rho^0&=&  e c_i^0 \exp\left[- \gamma (\rho_{\rm ext}^0+\rho^0)\right]
-e c_i^0 \exp\left[\gamma (\rho_{\rm ext}^0+\rho^0)\right]
\nonumber\\
&=& - 2 e c_i^0 \sinh\left[\gamma (\rho_{\rm ext}^0+\rho^0)\right]. 
\nonumber\\
&\approx& - 2 e c_i^0 \sinh \left[\gamma^{\rm apprx} (\rho_{\rm ext}^{0,{\rm apprx}} +\rho^0)\right].
\end{eqnarray}
We note that $\gamma^{\rm apprx}$ has dimensions of inverse charge density, so it is instructive to study the dimensionless charge density,  $\tilde{\rho}=\gamma^{\rm apprx} \rho^0$. According to Eq.~\ref{gammaDef}, a dimensionless charge density with a value of $1$ describes a system where the distance between ions is proportional to the Bjerrum length. Hence, we expect strong ion-ion interaction when the dimensionless charge density is large, while for weak charge densities ($\tilde{\rho} \ll 1$), we expect weak electrostatic interactions, that would result in the breakdown of electroneutrality. 

In terms of the dimensionless charge density we obtain the following algebraic equation: 
\begin{eqnarray}
\label{EqRhoSinh}
 \tilde{\rho} &=& -\xi\sinh(\tilde{\rho}_{\rm ext}+\tilde{\rho}),
\end{eqnarray}
where the two parameters, $\xi$ and $\tilde{\rho}_{\rm ext}$, are related to two important length-scales of the system, the Debye-length and the Gouy-Chapman length:
\begin{eqnarray}
 \tilde{\rho}_{\rm ext}&=&  \gamma^{\rm apprx} \rho_{\rm ext}^{0,{\rm apprx}} = \frac{2R}{\lambda_{\rm GC}} \log\left(\frac{L}{2R}\right) 
 \nonumber\\
  \xi&=&2ec_0 \gamma^{\rm apprx} = \frac{1}{2}\left(\frac{R}{\lambda_D}\right)^2\log\left(\frac{L}{2R}\right).
\end{eqnarray}
The Gouy-Chapman length ($\lambda_{\rm GC}=e/2\pi q_s  \lb$) characterizes the distance from a charged surface at which its electrostatic and thermal energies are equal. If this distance is much smaller than the pore radius, the system is effectively three-dimensional, and electroneutrality holds. As pore charge decreases, $\lambda_{\rm GC}$ becomes comparable with the pore radius and the one-dimensional geometry is recovered. Our parameter space is therefore described by three length-scales: the Debye screening length, Gouy-Chapman length, and the pore diameter. The breakdown of electroneutrality into a disordered phase is further augmented by the aspect ratio of the system, which effectively re-scales the pore diameter. 

The solution to Eq. \ref{EqRhoSinh} can be written as $\tilde{\rho}=-\tilde{\rho}_{\rm ext}+f_\xi^{-1}(\tilde{\rho}_{\rm ext})$, where $f_\xi(x)=x+\xi\sinh(x)$. Approximated solutions can be found for highly and weakly charged pores:
\begin{eqnarray}
\label{rhoApprxSol}
\tilde{\rho}= f_\xi^{-1}(\tilde{\rho}_{\rm ext})-\tilde{\rho}_{\rm ext} = \begin{cases}
    -\frac{\tilde{\rho}_{\rm ext}}{1+\xi^{-1}}, & \tilde{\rho}_{\rm ext} \ll 1.\\
     \log\left(\frac{\tilde{\rho}_{\rm ext}}{\xi}\right)-\tilde{\rho}_{\rm ext} . & \tilde{\rho}_{\rm ext}\gg 1.
  \end{cases}\nonumber\\
\end{eqnarray}
This solution is illustrated in Fig.~\ref{fig_UEPpd} as a function of $\lambda_{GC}$ and $\lambda_D$, for a pore with dimensions $R=1$nm and $L=100$nm.  We identify four different regimes, as shown in the four panels of Fig.~\ref{fig_UEPpd}. At low surface charge (large $\lambda_{GC}$) and short screening lengths (thin double layer), the system can be described by the standard DH approximation, and the electroneutrality assumption is valid. This theory also covers the beginning thick double layer regime, where the double layers begin to overlap. For high surface charges the linearized Poisson-Boltzmann equation is no longer valid. In this strong coupling regime the full non-linear equation is required, but interestingly, it also ensures the ions will completely screen any surface charge. Our solution, however, predicts a fourth regime, of low surface charge and small concentration. Under these conditions, electroneutrality is broken, and external fields must be accounted for. 

Two theoretical curves mark the boundaries of the electroneutral phase. The vertical line is derived from the weak charge approximation, as the Debye length at half screening:
\begin{eqnarray}
\label{vPhaseDiagramBoundary}
\lambda_D = R\sqrt{2\log \frac{L}{2R}} .
\end{eqnarray}
The second curve mark the transition to electruneutrality due to high surface charge, and is obtained by requaring half screening in the strong coupling limit of Eq.~\ref{rhoApprxSol}:
\begin{equation}
\label{phaseDiagramBoundary}
    \lambda_D=\sqrt{\lambda_{\rm GC} R}\left(\frac{L}{2R}\right)^{R/2\lambda_{\rm GC}}, 
\end{equation}
which asymptotes to a horizontal line at $\lambda_{GC}\approx R$ as the concentration decreases. Note that the transition to electroneutrality is slow, and spans roughly an order of magnitude change in the parameters. Electroneutrality breakdown can thus play a major role in the physics of nanopores.

\begin{figure}[htp]
  \includegraphics[width=\columnwidth ]{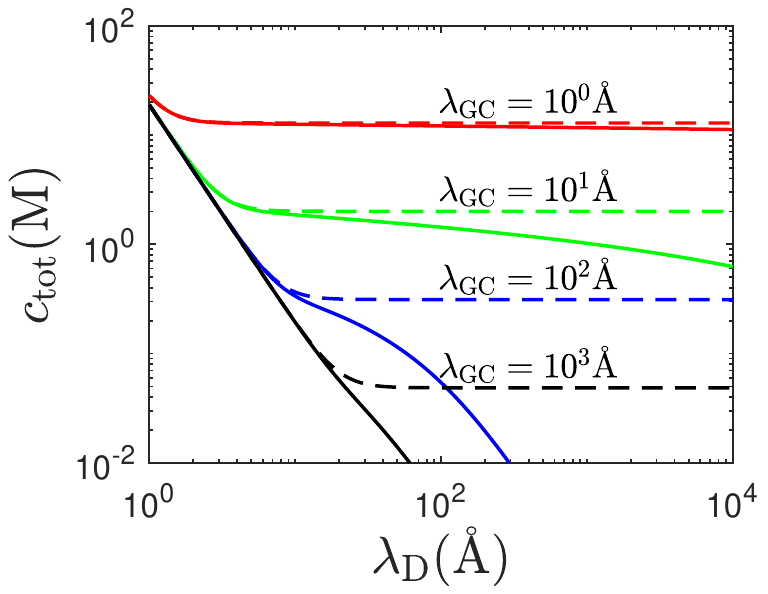}
  \caption{The total ionic concentration in a nanotube, as a function of concentration for different surface charges. Dashed lines show concentration calculated for electroneutral pores. }
  \label{fig_ionicConcentration}
\end{figure}

The total ionic concentration inside the nanopore, depicted in Fig.~\ref{fig_ionicConcentration}, is related to the pore charge by Donnan equilibrium:
\begin{equation}
\label{totalConcentration}
 c_{\rm tot} = \sqrt{(2 c_0)^2 + (\rho^0/e)^2}.
\end{equation}
In electroneutral systems the concentration reaches a plateau in the dilute limit, where the only ions inside the channel balance the surface charge. Accounting for electroneutrality breakdown, however, significantly alters the behavior in the dilute limit. 

\subsection{Sub-nanometer nanopores: dehydration and images forces}
Ion specific effects have important consequences on the behavior of nanopores, especially in the sub-nanometer scale. In extremely narrow pores ions have to remove their hydration shell, which creates a large energy barrier for entering the pore \cite{epsztein2019activation,sahu2017ionic}. Denoting this energetic cost for the $i$th specie by $E_i$, we generalize the self-consistent equation for the mean charge distribution:
\begin{eqnarray}
\label{rhoEqn}
\rho^0=\sum_{i=1}^N e z_i c_i^0 \exp\left[-\beta E_i - z_i \gamma (\rho^0+\rho_{\rm ext}^0)\right].
\end{eqnarray}
In the monovalent case the pore-charge equation takes a similar form to Eq. (\ref{EqRhoSinh}), with re-scaled coupling parameter $\xi$ and external charge $\tilde{\rho}_{\rm ext}$:
\begin{eqnarray}
\label{EqRhoSinh2}
 \tilde{\rho}_{\rm ext}&=&  \frac{2R}{\lambda_{\rm GC}} \log\left(\frac{L}{2R}\right) + \frac{E_+ - E_-}{2\kbt}
 \nonumber\\
  \xi&=&2\left(\frac{R}{\lambda_D}\right)^2\log\left(\frac{L}{2R}\right)\exp\left(-\frac{E_++E_-}{2\kbt}\right).
\end{eqnarray}
Hence, the phase diagram (Fig.~\ref{fig_UEPpd}) remains similar, but skewed: the x-axis is rescaled by the average Boltzmann weight, while the y-axis is shifted by dimensionless energy difference. Any asymmetry in the dehydration energy will result in an excess charge within the pore since the dehydration energy plays a similar role to that of the surface charge. 

Even though we incorporate in our model an energy barrier, it is important to emphasize again that we are considering a model with a constant permittivity everywhere. The differences in self-energies are only one aspect of a dielectric mismatch, that can change ion-ion and ion-wall interactions as well\cite{kamenev2006transport,zhang2005conductance,zhang2006ion, levin2006electrostatics,bordin2012ion}.

\subsection{Towards a general theory: the electric leakage boundary conditions and validation the UEP model}
We examined in this section a quantitative model for electroneutrality breakdown, and obtained a simple algebraic equation that determines the net charge inside a nanopore. In agreement with the numerical analysis of section II, we find that breakdown occurs only in the thick doubly layer limit. Yet, our model does not allow us to resolve the spatial distribution of ions inside the channel. As we show in section II, the full charge density can be found by solving the combination of Laplace and PB equations (Eq.~\ref{PBL_eqns}) with the BC of Eq.~\ref{PBL_BC}. 

For narrow and long pores ($L\gg R$) it is tempting to simply take $L\rightarrow\infty$ and solve a $1$d ordinary differential equation. However, as we have shown here, even in very long channels the lengths plays an important role. Hence, solving a cumbersome set of two coupled nonlinear $2d$ PDEs to get the full charge density is required. For weakly charged surfaces (DH regime) an exact analytical solution is available in the form of an infinite series (see Appendix A), but it is difficult to gain any insight from this complicated expression.

It turns out, however, that for long and narrow channels we can go one step beyond the UEP model, and obtain approximated boundary conditions for PB equation. With this approximated BC, we can again solve a simple $1$d ODE, but a one that inherently captures electroneutrality breakdown. 

Recall that the in cylindrical coordinates boundary conditions are:
\begin{equation}
    \varepsilon_{\rm in} \frac{\partial \phi_{\rm in} (r,z)}{\partial r} = \varepsilon_{\rm out} \frac{\partial \phi_{\rm out} (r,z)}{\partial r} + q_s
\end{equation}

In Appendix B we show that the electric field and electrostatic potential in the outer region are related by a simple expression:
\begin{equation}
\label{bc_out}
    \partial_r\phi_{\rm out}(R,z) \approx  -\frac{\phi_{\rm out}(R,z)}{M_{L/R} R}. 
\end{equation}
where $M_{L/R}$ is defined as:
\begin{equation}
    M_{L/R} = \log\left(\frac{2 L}{\pi R}\right)- \gamma_{\rm Euler},
\end{equation}
and $\gamma_{\rm Euler}\approx0.577$ is Euler's constant. Using the electrostatic potential continuity condition, the new Robin type ``electric leakage" boundary condition is:
\begin{equation}
\label{leakageBC}
    \partial_r\phi_{\rm in}(R)=\frac{q_s}{\varepsilon_{\rm in}}- \frac{\varepsilon_{\rm out}}{\varepsilon_{\rm in} R M_{L/R} } \phi_{\rm in}(R),
\end{equation}

\begin{figure}[htp]
  \includegraphics[width=\columnwidth ]{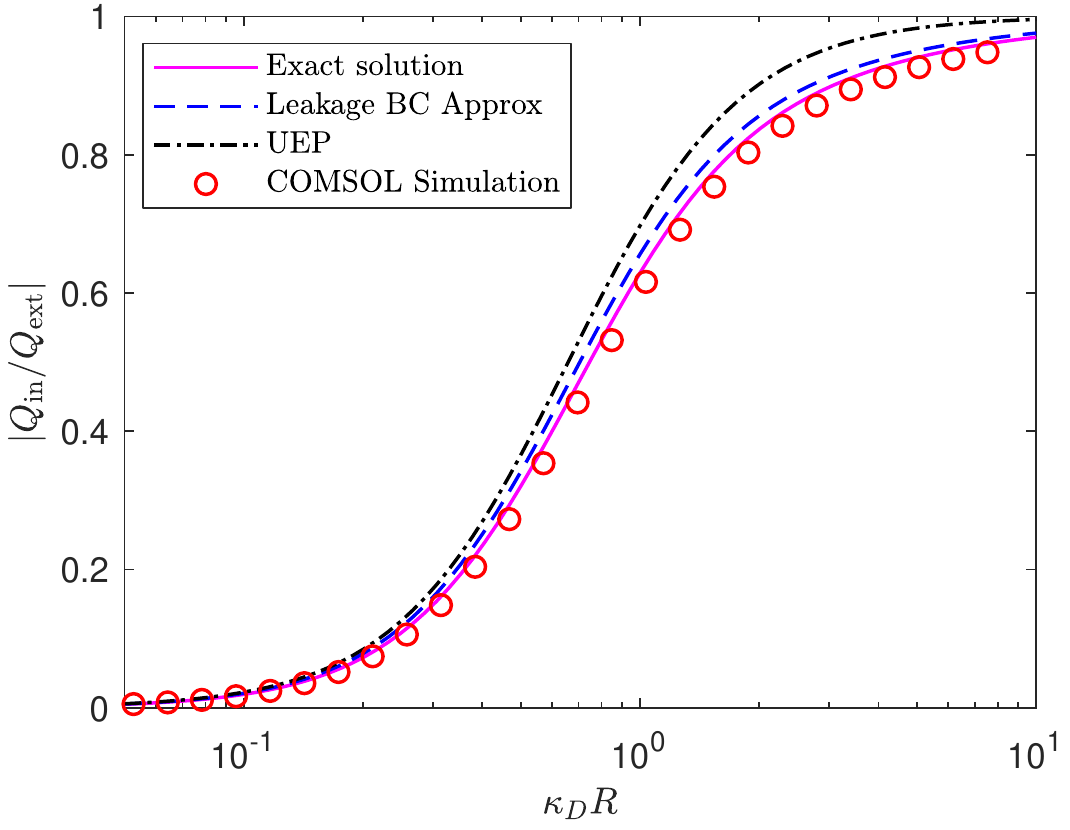}
  \caption{Charge accumulated in a weakly charged pore as a function of inverse Debye length, evaluated based on exact analytical solution (solid line), numerical solution of the PB equation with the Robin boundary condition of Eq.~\ref{leakageBC} (dashed line), and the UEP model (dash-dotted line, Eq.~\ref{EqRhoSinh}. Results are compared with COMSOL simulation (circles). The pore dimensions are: $R=5$nm and $L=1\mu$m. }
  \label{fig_exact_BC_uep}
\end{figure}

In Fig.~\ref{fig_exact_BC_uep} we compare a numerical simulation to the exact analytical solution (Eq.~\ref{exactSol}) for the total charge in a weakly charged nanopore, as well as to the UEP model and the new BC. All approximated models are compared with the numerical COMSOL simulations as well. To satisfy the weak potential assumption ($e \beta \phi \ll 1$) of the Debye-H\"uckel approximation, we focused only on the limit where the charge density approaches $0$. As expected, in the dilute limit the UEP model, as well as the approximated boundary condition, perfectly match the exact analytical result. As concentration increases, the uniform approximation assumption is no longer valid. 


The excellent match of the tractable UEP model to more elaborate schemes allows us to use it when we are interested in fitting experimental measurements. In section VII we will explore some available data, and show the importance of electroneutrality breakdown to transport properties. First, however, we would like to take a small detour and discuss {\it why} electroneutrality was obtained. As we mentioned in the introduction, we claim that this is a unique feature of the $1$d geometry. In the next two sections we discuss how the basic physics of $3$d electrostatic interactions between particles confined to a $1$d narrow channel, and show that the UEP model is best understood as an example of $1$d theory. The impatient reader, however, can skip directly to section VI. 



\section{Screening length for ions confined to lower dimensions}
A central ion in an aqueous solution is surrounded by a screening cloud meaning that ions in its vicinity tend to be of opposite charge. In three dimensions, we can find an analytical expression for the shape of the screening cloud, achieved by solving the linearized Poisson-Boltzmann equation. The density profile of screening ions decays exponentially with a characteristic length, the ``Debye length" ($\ld$). When ions are restricted to different dimensionality, still interacting with a $1/r$ pair-wise potential, there is no similarly tractable equation to determine the shape and size of the screening cloud. We will show in the following sections, that an equivalent screening length exists and plays an important role in the physics of lower dimension electrolytes, but it formally requires a cumbersome derivation. Before we delve into a more rigorous formalism, we first present a simple scaling argument for the screening length that holds in any dimension. 

The screening length, in essence, is the distance at which entropic and electrostatic forces balance each other. The electrostatic force pulls the screening cloud closer to the central ion, while entropy favors uniform charge distributions and pushes the screening cloud away. If we consider a screening cloud of radius $\lambda_s$ with $N_s$ ions around a negatively charged univalent ion, the probability of each ion to have charge $+e$ equals $p^+_i = 1/2+1/2N_s$.  Maximal entropy is achieved if there is an equal probability for an ion to be positive or negative. By forcing the ions to screen the central charge, the probability of positively charged ions slightly increases, and the entropy is reduced. Note that the ions outside the screening cloud have equal probabilities to have positive/negative charge. The entropy associated with this screening cloud equals (assuming $N_s\rightarrow\infty$, which corresponds to the dilute limit):
\begin{eqnarray}
S &=& -k_{\rm B} \sum_{i=1}^N\left[p_i^+ \log p_i^+ + p_i^-\log p_i^- \right] 
\nonumber\\
& = & -N_s k_{\rm B}\left[\left(\frac{N_s+1}{2N_s}\right)\log\left(\frac{N_s+1}{2N_s}\right)\right.
\nonumber\\
&+&\left.\left(\frac{N_s-1}{2N_s}\right)\log\left(\frac{N_s-1}{2N_s}\right)\right] + (N-N_s) k_{\rm B} \log 2
\nonumber\\
&\approx&  k_{\rm B}\left(N\log2 -\frac{1}{2 N_s}\right),
\end{eqnarray}
where $N$ is the total number of ions in the system, both inside and outside the screening cloud. The first term is constant (does not depend on $N_s$) and can be ignored. Note that this is only the entropy associated with the possibility of each ion to be positive or negative, and positional entropy is neglected.  Relating the entropy to the $d$-dimensional sphere volume\cite{mclaren1958introduction}, $V=A_d \lambda_s^d/d$, where $A_d$ is the surface area of a unit sphere, and average ionic pair concentration ($2c=N_s/V$), we find that the entropy of ions inside the sphere is reduced by:
\begin{eqnarray}
\Delta S =  - \frac{d k_{\rm B} }{4A_d  c \lambda_s^{d}}.
\end{eqnarray}
The electrostatic energy of a uniformly distributed screening cloud equals:
\begin{eqnarray}
U&=& -\frac{e^2A_d}{4\pi \varepsilon V }\int_a^{\lambda_s}r^{d-2} {\rm d}r 
\nonumber\\
&=& -\frac{e^2 }{4\pi \varepsilon \lambda_s}\times\begin{cases}
    \frac{d}{d-1}, & d>1.\\
    \log\frac{\lambda_s}{a}, & d=1,
  \end{cases}
\end{eqnarray}
where $a$ is the ion size, assumed to be much smaller than $\lambda_s$.
 Here, we make the critical assumption of a $3$d Coulomb potential decaying as $1/r$, which effectively spills out of the confining geometry, and neglect for now any dielectric response of the surrounding matrix, which modifies the result but does not alter the basic scaling arguments (as shown below).

\begin{figure}[htp]
  \includegraphics[width=\columnwidth ]{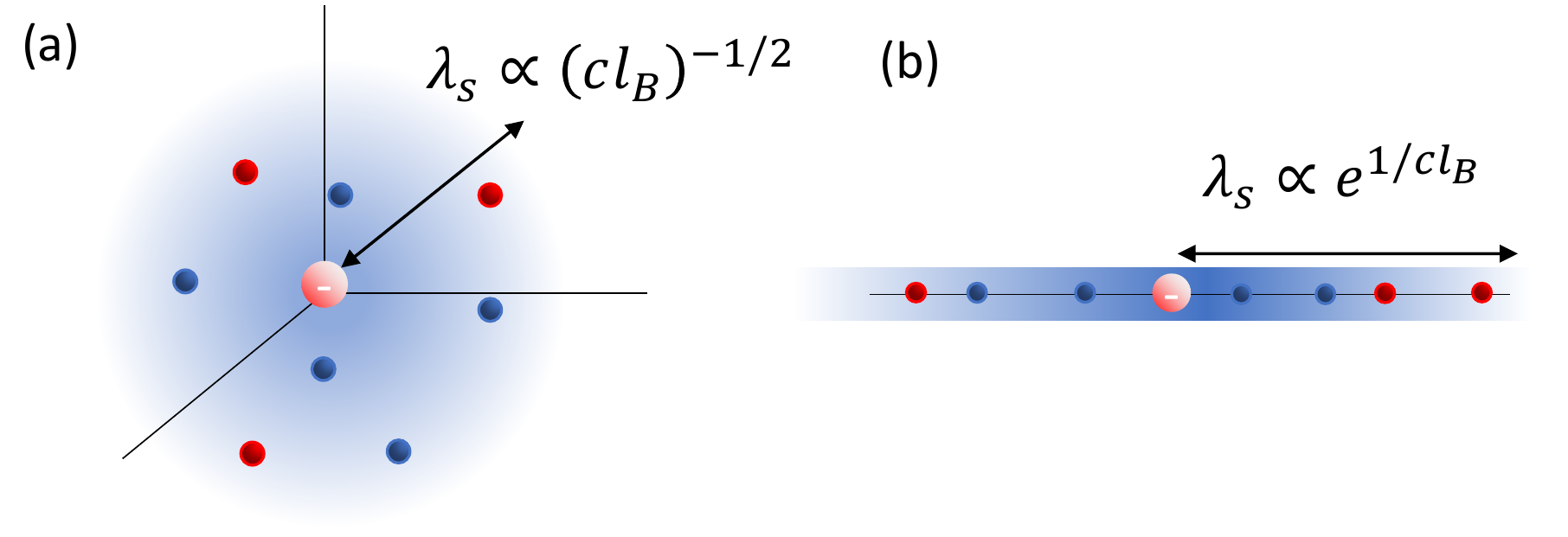}
  \caption{Illustration of the standard Debye screening in 3d(a), and the extended screening length of 1d(b), according to  Eq.~\ref{ScalingLambda}.}
  \label{fig_scaling}
\end{figure}

Minimizing the free energy ($F=U-T\Delta S$)  with respect to $\lambda_s$ , and ignoring the numerical prefactors,  yields the following scaling behavior for the screening length:
\begin{eqnarray}
\label{ScalingLambda}
\lambda_s \propto \begin{cases}
    (c \lb)^\frac{1}{1-d}, & d>1\\
    a\exp(\frac{1}{8c\lb}), & d=1,
  \end{cases}
\end{eqnarray}
where $\lb = e^2/4\pi \kbt \varepsilon$ is the Bjerrum length. In three dimensions, we recover the standard Debye screening: $\lambda_s\propto\lambda_D =  (8\pi c l_B)^{-1/2}$. In a nanoslit geometry of two dimensions, the screening length is proportional to $(c l_B)^{-1}$. Forcing ions to reside on a two-dimensional plane reduces electrostatic interactions, and slightly increases the size of the screening clouds. A much more dramatic effect is observed  for one-dimensional confinement in a long, thin nanopore, where the screening length is {\it exponentially} large, as shown in Fig.~\ref{fig_scaling}. 

 If we consider our one-dimensional system to be a cylinder with radius $R$, length $L$, and a $3$d concentration (per volume) of  $c_0$, the corresponding $1$d concentration (per length) in the axial direction is $c^{\rm 1d} = \pi R^2 c_0$. In terms of these bulk properties, we find that the dimensionless factor, $c^{\rm 1d} l_B= \pi R^2  c_0l_B$, is related to the ratio of Debye length to pore radius, which enters the exponential scaling of the screening length in one dimension: $\lambda_s \propto a\exp\left[(\ld/ R)^2\right]$. For narrow pores in dilute electrolytes, the Debye length can easily be greater than the pore radius,  which is the traditional limit of ``thick double layers" ($\lambda_D \gg R$), but in contrast to classical continuum models, we predict that this results in an extremely large screening length, easily exceeding all geometrical scales in the problem.  In particular, when the screening length exceeds the total length of the channel($\lambda_s \gg L$), the central ion is no longer screened, and electroneutrality breaks down.  

\section{Mean-field theory on a one-dimensional lattice}

 In the next section we compare the UEP model with experimental data, and show it can be used to interpret conductance curves. Before doing so, we would first like to show quantitatively how electroneutrality breakdown is a fundamental property of $1$d confinement, by considering a system of ions restricted to reside along a line. This will also allow us to explore ion-ion correlations along the pore axis, and observe a transition from order to disorder.  

We study a lattice-gas model, and not a continuum model, for two reasons. First, it will enable us to discuss packing constraints at the high concentration limit. A more fundamental reason was hinted in the previous section: there is no equivalent continuum PB model in $1$d. The scaling argument showed that the screening length depends on the minimal distance between ions ($a$), and this will remain valid in the analysis here as well. As a result, we cannot find a corresponding differential equation that describes the system in the continuum limit.  Note that in contrast to many previous $1$d models of electrolytes (\cite{edwards1962exact,lenard1961exact,demery2012overscreening}), the electrostatic interaction is three dimensional ($1/r$): we study point-like ions along a line, and not parallel charged sheets. 

\begin{figure*}[!]
  \includegraphics[width=\textwidth ]{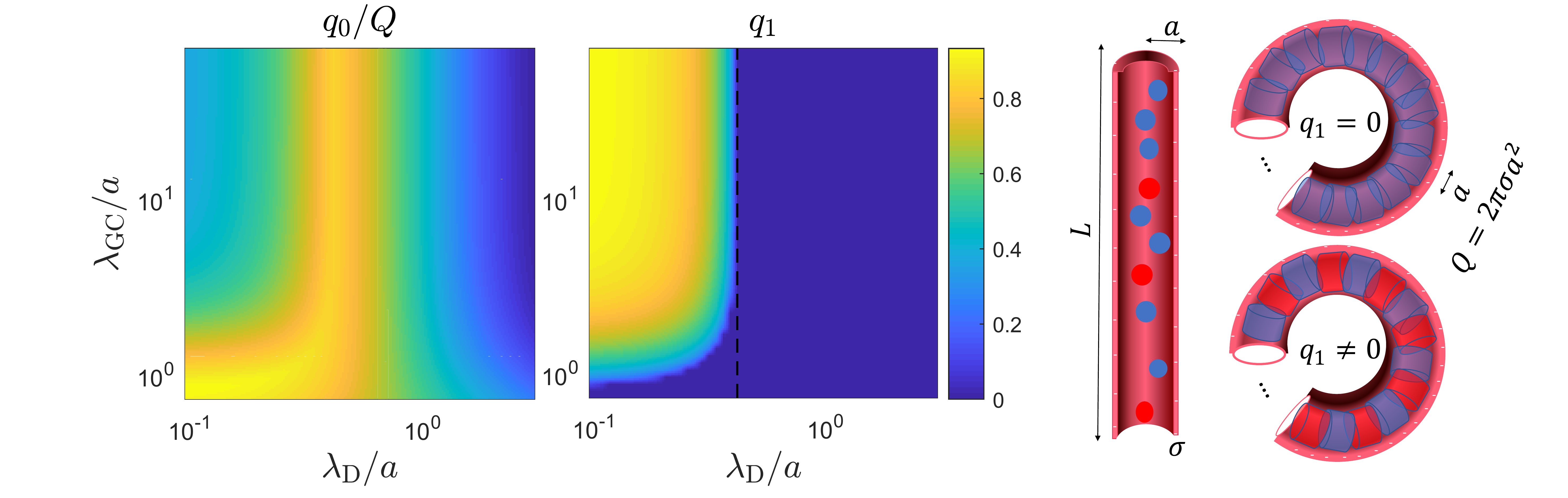}
  \caption{Solution of the 1D-ring equation with uniform surface charge (Eq.~\ref{mpb_2}). The solution, $q_i=q_0+q_1(-1)^i$, has an average part ($q_0$, left figure) that screens the external charge ($Q$) and an oscillatory part ($q_1$, right figure). Solutions are presented as a function of the bulk Debye length ($\lambda_D$) and the Gouy-Chapman length ($\lambda_{\rm GC}$). The dashed vertical line is the critical Debye length, calculated by Eq.~\ref{criticalFug}.}
  \label{fig_1dPD}
\end{figure*}

For mathematical convenience, we consider the free energy functional of a 1d periodic lattice model (a ring), with lattice spacing equal to the ionic size, $a$ (see Fig.~\ref{fig_1dPD}). The $i^{\rm th}$ site can be occupied by a positive ion, negative ion or a vacancy, with probabilities $p_i^+$, $p_i^-$ and $1-p_i^--p_i^+$, respectively. The electrostatic energy, $U$, is given by:

\begin{eqnarray}
    U&=&\frac{1}{8\pi}\sum_{i \neq j}\frac{e^2 (p^+_i-p^-_i)(p^+_j-p^-_j)}{a \varepsilon|i-j|}
    \nonumber\\
    &+& \sum_i e (p^+_i-p^-_i) \phi^{\rm ext}_i,
\end{eqnarray}

where $\phi^{\rm ext}_i$ is an external electrostatic potential. We denote the dimensionless average charge vector $q_i=(p^+_i-p^-_i)$ and the dimensionless interaction matrix $\tilde{\Phi}_{ij}=|i-j|^{-1}$, so the electrostatic energy reads: 
\begin{equation}
\label{Udef}
U=\frac{e^2}{8\pi a \varepsilon}\left(\qbf^T \tilde\Phi\qbf\right) + e\qbf^T \phi^{\rm ext}.    
\end{equation}
To calculate the free energy, we add the entropy of mixing:
\begin{eqnarray}
 S&=& -k_{\rm B} \sum_i \left[p^+_i\log p^+_i  + p^-_i\log p^-_i +\right.
 \nonumber\\
 & &\left.(1-p^+_i-p_i^-)\log (1-p^+_i-p_i^-)\right] .  
\end{eqnarray}
If connected to a particle reservoir, the chemical potential is set externally and is calculated by the functional derivative of the free energy density $f=(U-TS)/a$ with respect to concentration ($c^\pm_i=p^\pm_i/a$):
\begin{eqnarray}
\mu_\pm = \kbt \log\left(\frac{\pbf_\pm}{1-\pbf_+-\pbf_-}\right)\pm \frac{e^2 \tilde{\Phi}\qbf}{4\pi a \varepsilon}\pm e\phi^{\rm ext}.\nonumber\\
\end{eqnarray}
Rearranging the terms, we obtain the $1$d analog of the Bikerman model (\cite{Bikerman1942, Borukhov1997, kornyshev2007double,kilic2007steric,bazant2009towards}):
\begin{eqnarray}
\label{mpb}
\qbf = -\frac{2\Lambda \sinh \left[\frac{l_B}{a} \tilde{\Phi}\qbf+e\beta \phi^{\rm ext} \right]}{1+2\Lambda \cosh\left[\frac{l_B}{a} \tilde{\Phi}\qbf+e\beta \phi^{\rm ext} \right] },
\end{eqnarray}
where the fugacity $\Lambda={\rm e}^{\beta \mu}$ is proportional to the bulk reservoir ion activity. 

The general form of the $1$d ring equation is not very illuminating, and specific examples are required to show how ion-ion correlations play an essential role in this model. We study two systems: a charged homogeneous nanopore and the charge distribution around a central ion. In the dilute limit, we recover the behavior described in the previous section. As ionic concentration increases, the model naturally predicts a transition to an ordered structure, including the short-range over-screening phenomenon in intermediate concentrations.  
\begin{figure}[htp]
  \includegraphics[width=\columnwidth ]{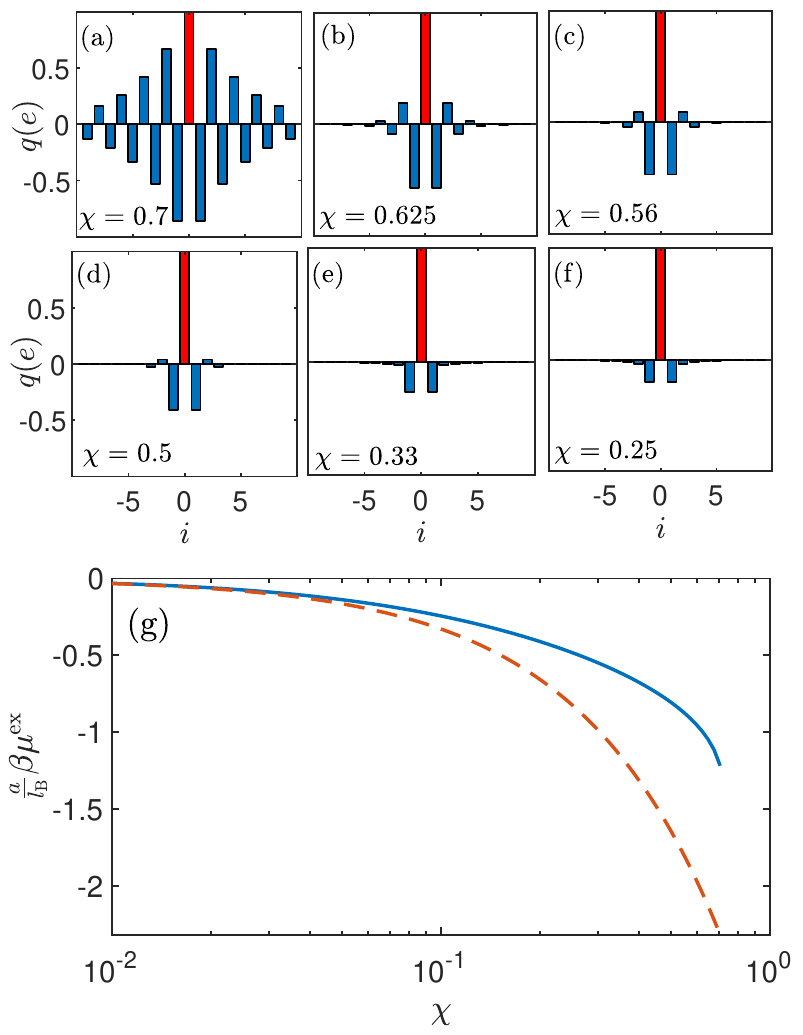}
  \caption{Charge distribution around a central ion and the resulting activity coefficient, for different concentrations. (a-f) Charge distribution around a central ion for six different coupling strengths ($\chi$). For weak coupling , the central ion is screened only by oppositely charged ions (e-f). As the coupling increases, over screening and oscillations are observed. (g) Ionic activity in a nanopore, based on the exact 1D solution for the mean field equation (Eq.~\ref{fullActivity}, solid blue line) and the dilute limit approximation (Eq.~\ref{approxActivity}, dashed red line).} 
  \label{fig_1dActivity}
\end{figure}
\subsection{A Uniformly Charged 1-d Nanopore}
Let us assume we have a homogeneous charge density in our system: $q_i=q$, $\phi^{\rm ext}_i = \phi^{\rm ext}$. For periodic boundary conditions, a uniform charge distribution is an eigenvector of the interaction matrix $\tilde{\Phi}$, where the eigenvalue is twice the harmonic number $H_{L/a}$. For a long chain ($L\gg a$) the harmonic number can be expanded:
\begin{eqnarray}
H_{L/a}\approx \log\left(\frac{L}{a}\right)+\gamma_{\rm Euler}+\frac{a}{2L}+\ldots,
\end{eqnarray}
If the external potential is due to a uniform charge distribution on the pore walls ($Q$ per site), Eq. \ref{mpb} becomes the following algebraic equation: 
\begin{eqnarray}
\label{mpb_1}
q = -\frac{2\Lambda \sinh \left[2\frac{l_B}{a}H_{\frac{L}{a}} \left(q+Q\right) \right]}{1+2\Lambda \cosh \left[2\frac{l_B}{a}H_{\frac{L}{a}} \left(q +Q\right)\right] },
\end{eqnarray}
In the dilute limit, this equation has a similar form to Eq.~\ref{EqRhoSinh}. As the external charge increases, a maximal charge density of one charge per lattice site prevents an unphysical accumulation of charges inside the pore. 

Limiting the $1$d-ring equation to only uniform distributions explicitly neglects any ion-ion correlations. When ion-ion interactions are strong enough, the system will form a crystal structure. In contrast to the $3$-dimensional case, where the PB equation cannot predict ordered structures, our $1$d model can be easily extended to include the expected phase transition. To  account for patterns of alternating signs, we generalize our argument and replace the constant charge density with the form: $q_i = q_0 + q_1 (-1)^i$. Both uniform charge density and an alternating pattern are eigenvectors of the interaction matrix $\tilde{\Phi}$, where the eigenvalue of $(-1)^i$ equals $-2\log2$. With this functional form we obtain a set of coupled non-linear algebraic equations, for the average charge at even/odd locations: 
\begin{eqnarray}
\label{mpb_2}
q_0 - q_1 &=& -\frac{2\Lambda \sinh \left[-2q_1\frac{l_B}{a}\log(2) + 2\frac{l_B}{a}H_{\frac{L}{a}} \left(q_0+Q\right) \right]}{1+2\Lambda \cosh \left[2q_1\frac{l_B}{a}\log(2) +2\frac{l_B}{a}H_{\frac{L}{a}} \left(q_0 +Q\right)\right] }
\nonumber\\
q_0 + q_1 &=& -\frac{2\Lambda \sinh \left[2q_1\frac{l_B}{a}\log(2) +2\frac{l_B}{a}H_{\frac{L}{a}} \left(q_0+Q\right) \right]}{1+2\Lambda \cosh \left[-2q_1\frac{l_B}{a}\log(2) +2\frac{l_B}{a}H_{\frac{L}{a}} \left(q_0 +Q\right)\right] },
\nonumber\\
\end{eqnarray}

Exploring the space of solutions of Eq.~\ref{mpb_2} is shown as a phase diagram in Fig.~\ref{fig_1dPD}, for $a=5$\AA  and $L=100$\AA. As long as the ion-ion correlations are weak, we recover the same behavior found in the continuum model, including electroneutrality breakdown in the dilute limit. However, we find another breakdown of electroneutrality in the high concentration limit. This is the oscillatory regime ($q_1>0$). Since $q_1+q_0<1$, an increase of oscillating term, $q_1$, has to come at the expense of the average term, $q_0$, and electroneutrality is broken. With an increase of external charge, oscillations are suppressed and electroneutrality is again favored.
As shown in Fig.~\ref{fig_1dPD}, the order-disorder phase boundary only weakly depends on the surface charge and can be evaluated analytically based on the $Q=0$ limit. In this limit the average charge is $q_0=0$, and the number of solutions is determined by a single algebraic equation:
\begin{eqnarray}
\label{mpb_3}
q_1 &=& \frac{2\Lambda \sinh \left(2q_1\frac{l_B}{a}\log2 \right)}{1+2\Lambda \cosh \left(2q_1\frac{l_B}{a}\log2 \right)}
\nonumber\\
\end{eqnarray}
The RHS of Eq.~(\ref{mpb_3}) is monotonically increasing, starting from $0$, and has a monotonically decreasing slope. Hence, a second solution is available only if the slope at $q_1=0$ is greater than $1$, which leads to a critical fugacity of:
\begin{equation}
\label{criticalFug}
    \Lambda_{\rm cr} = \frac{a}{4l_B \log 2  -2a}.
\end{equation}
The results in Fig.~\ref{fig_1dPD} are displayed in terms of the standard $3$d system parameters. With a pore radius $a$, the surface charge density equals $q_s=Q/(2\pi a^2)$. The bulk ionic concentration is related to the fugacity $c^{\rm 3D} = \Lambda/\pi a^3$. The Gouy-Chapman length and the Debye length are defined as usual.

\subsection{Charge distribution around a central ion}
Bulk oscillations are maintained only for concentrations beyond a critical concentration, with persisting long-range order. Slightly below the critical density, we expect temporary short ranged decaying oscillations, that will eventually be replaced by monotonic decaying fluctuations in the dilute limit. We show how this behavior emerges with a standard Debye-H\"uckel approach, by solving for the charge distribution around a central ion. 

We take advantage of the periodic boundary conditions, and constrain the $i=0$ site to have a charge $Qe$ by adding a term $\alpha e^2/a\varepsilon(q_0-Q)$ to the free energy functional (Eq.~\ref{Udef}), where $\alpha e^2/a\varepsilon$ is a Lagrange multiplier. The resulting mean-field equation reads:
\begin{eqnarray}
\qbf = -\frac{2\Lambda \sinh \left[\frac{l_B}{a}\left( \tilde{\Phi}\qbf+\alpha\delta_{i,0} \right)\right]}{1+2\Lambda \cosh \left[\frac{l_B}{a} \left(\tilde{\Phi}\qbf+\alpha\delta_{i,0}\right)\right] },
\end{eqnarray}
which we solve in the linear Debye-H\"uckel regime:
\begin{eqnarray}
q_i & = & -\alpha\left(\tilde{\Phi}+\chi^{-1}\right)_{i,j}^{-1}\delta_{j,0},
\end{eqnarray}
where the coupling parameter $\chi=\frac{2\Lambda l_B}{a}$ is defined as the ratio of ionic spacing ($a/2\Lambda$) and the Bjerrum length. By the translation symmetry of the matrix $\tilde{\Phi}_{ij}=\tilde{\Phi}_{|i-j|}$, we find a closed-form expression in the discrete Fourier space:
\begin{eqnarray}
\label{1dpb_sol}
q_i &=&  \frac{\alpha}{4\pi}\int_{2\pi}{\rm d}\omega \frac{\cos(\omega i) }{2\log\left|2\sin\frac{\omega}{2}\right|-\chi^{-1}}, 
\end{eqnarray}
where the normalization constant $\alpha$ is set such that $q_0=Q$. Note that the solution is only valid in the disordered phase, where $\chi^{-1} > 2\log 2$, which coincides with the critical fugacity (Eq.~\ref{criticalFug}) in uniformly charged nanopores if size effects are neglected ($a\ll \lb$). Fig.~\ref{fig_1dActivity} shows charge density profiles for different concentration, illustrating how Eq.~\ref{1dpb_sol} is able to capture both the dilute Coulomb gas limit and the onset of long-range ordering, and predict the transition from screening, to over screening and oscillations. 
 
So far we assumed an infinitely long chain, that allowed us to get a closed-form solution in Fourier space (Eq.~\ref{1dpb_sol}). As a consequence, the electronutrality is guaranteed, and the total charge along the pore accumulates to $0$:
\begin{eqnarray}
\sum_{i=0}^\infty q_i = \alpha \lim_{\omega \rightarrow 0 }\frac{1}{2\log\left|2\sin\frac{\omega}{2}\right|-\chi^{-1}} \rightarrow 0.
\end{eqnarray}
However, the decay rate is {\it very} slow, and the screening cloud extends many lattice sites. If we look at $\tilde{q}(\omega)$ , the Fourier transform of $q_i$, we find a steep increase near $\omega=0$, on its way to its maximal value at $\omega=\pi$. We evaluate its width by finding the frequency at which $\tilde{q}(\omega)$ reaches half of its maximal height:
\begin{eqnarray}
\frac{1}{2\log\left|\Delta \omega \right|-\chi^{-1}}=\frac{1/2}{2\log 2-\chi^{-1}} \rightarrow \Delta \omega = 4 {\rm e}^{-\frac{a}{4\Lambda l_B}}.
\nonumber\\
\end{eqnarray}
Invoking the uncertainty principle, we conclude that the width of the screening cloud scales as ${\rm e}^{\frac{a}{4\Lambda \lb}}$, in agreement with our scaling analysis (Eq.~\ref{ScalingLambda}) and the electroneutrality condition (Eq.~\ref{ENcond}).  

Finally, we use our explicit solution of the screening cloud to evaluate the activity coefficient. For this purpose, the electroneutrality breakdown plays only a minor role. Ignoring the finite length of the system, the Debye-H\"uckel activity coefficient can be written as an integral expression:

\begin{eqnarray}
\label{fullActivity}
\beta \mu^{\rm ex} =  \frac{l_B}{a}\sum_i \frac{q_i}{i} = \frac{\alpha \lb }{2\pi a}\int_{2\pi}{\rm d}\omega \frac{\log\left|2\sin\frac{\omega}{2}\right|}{2\log\left|2\sin\frac{\omega}{2}\right|-\chi^{-1}},\nonumber\\
\end{eqnarray}
where $\mu^{\rm ex}$ is the excess chemical potential. In dilute systems we can expand the activity coefficient to lowest order in the coupling parameter $\chi$, and get:
\begin{equation}
\label{approxActivity}
    \beta \mu^{\rm ex} \approx - \frac{2\pi^2  l_B^2 }{3 a^2} \Lambda.
\end{equation}
The activity coefficient in confinement is much smaller than the bulk value. As ions cross over to the nanopore, they effectively shed off their ionic screening cloud. For a nanopore of radius $a$, with bulk ionic concentration ($c=\Lambda/\pi a^3$), the $1$d activity is only a fraction of the bulk one:
\begin{eqnarray}
\frac{\mu^{1D}}{\mu^{3D}}= \frac{\pi^2}{24}\frac{a}{\lambda_D},
\end{eqnarray}
where we used the standard Debye-H\"uckel activity coefficient, $\beta \mu^{3D}=l_B/2\lambda_D$. 

As shown in Fig.~\ref{fig_1dActivity}, this approximation is only valid in the dilute limit. As concentration increases, screening and over-screening dominates the electrostatic interactions, and reduce the activity coefficient further.

\begin{figure*}[!]
  \includegraphics[width=\linewidth ]{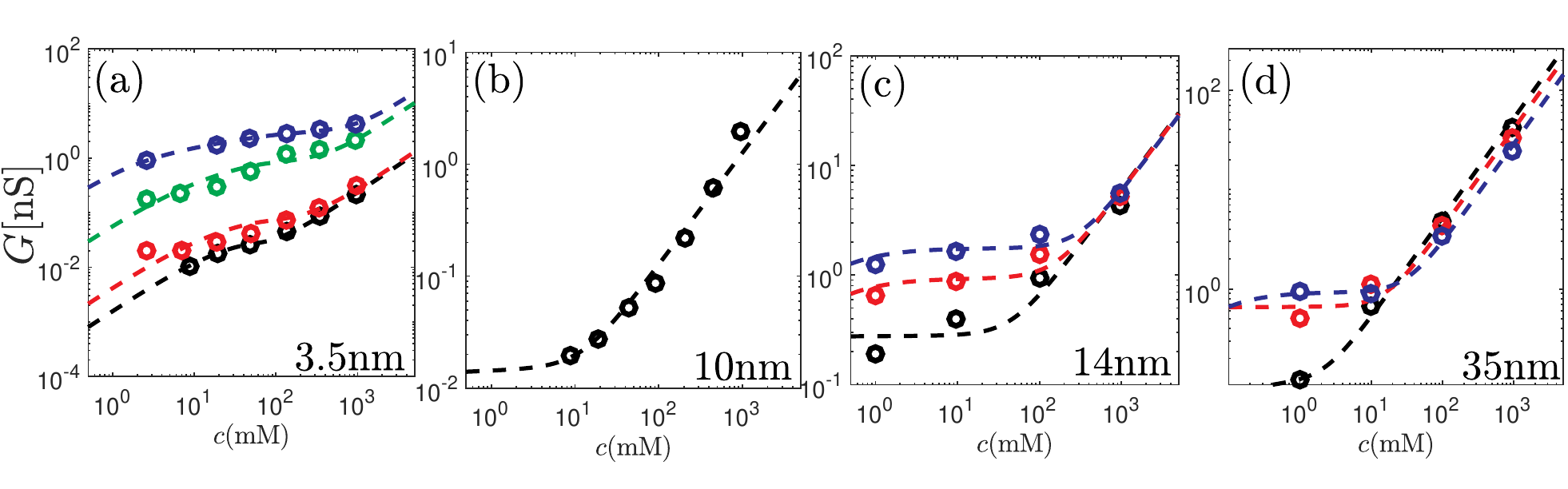}
  \caption{Conductance of a KCl solution as a function of concentration, inside CNT with varying surface charge and radii, fitted according to our model (dashed lines). The experimental data (circles) was adapted from \cite{Secchi2016},  where the surface charge was controlled by changing the pH . 
  (a) - A $3.5$nm wide pore, fitted barrier energy of $4.6\kbt$ and 4 different surface charges, from bottom to top: $-3\mCpA$ (black), $-1.6\mCpA$ (red), $3.4\mCpA$ (green), $5.2\mCpA$ (blue). (b) - A $10$nm wide pore, with only one surface charge: $12\mCpA$. (c) - A $14$nm wide pore, with $3$ different surface charge, fitted to: $63\mCpA$ (black), $213\mCpA$ (red) and $417\mCpA$ (blue). (d)- A $34$nm wide pore, with fitted surface charge of: $-7\mCpA$ (black), $55\mCpA$ (red) and $110\mCpA$ (blue). }
  \label{fig_Compare}
\end{figure*}
\section{Comparison with experiments}

\subsection{Single Digit Nanopores}


Our model predicts a non-trivial charge accumulation within the nanopore, due to the breakdown of electroneutrality, which has a direct implication on the measured conductivity of the pore. Assuming equal mobilities for all ions ($\upmu_D$), the conductance of the pore is given by:
\begin{eqnarray}
\label{eqnG}
G= e^2\upmu_D \frac{\pi R^2}{L} c, 
\end{eqnarray}
 where the concentration, $c$, is related to the accumulated charge by Eq.~\ref{totalConcentration}. We compare our results with conductance measurements in carbon nanotubes (CNT), taken from Ref.~\cite{Secchi2016}. It is important to note that the conductance behavior can be explained by different models. The CNT data was originally assumed to have a concentration-dependent surface charge and was later fitted by predicting the adsorption of hydroxyl ions to CNT pore walls (\cite{Biesheuvel2016}). Our goal is not to underestimate the importance of a charge regulation mechanism, that can lead to concentration-dependent surface charge by affecting the adsorption rate (\cite{dukhin2005electrokinetic}), but to suggest a plausible alternative with a constant surface charge. We show that the $1$d geometry by itself can lead to the variety of conductance curves observed in experiments. 

Fig.~\ref{fig_Compare} shows the conductance curves for CNTs with varying pore size and surface charge, as a function of KCl concentration. The experimental data were fitted according to Eq.~\ref{EqRhoSinh2}, with three fitting parameters- the ionic mobility, surface charge, and energy barrier. 

For the larger pore sizes, shown in Fig.~\ref{fig_Compare}(b-d), the energy barrier for entering the pore ($E_\pm$) was neglected, and the fitted ionic mobilities were on the order of the bulk mobility of KCl, and varied from $5-10\times10^{-8}{\rm m}^2/{\rm V s}$, compared to the bulk value of $7.62\times10^{-8}{\rm m}^2/{\rm V s}$ (see Fig.~\ref{fittedMobilities}, bulk value of ionic mobility was taken from \cite{lide2004crc}). This effectively only leaves one fitting parameter to determine the shape of the curves- the surface charge. As surface charge increases, electroneutrality is maintained especially for the larger pore radii. The curve approaches a constant in the dilute limit (see top curves in figures \ref{fig_Compare}c and \ref{fig_Compare}d). However, for smaller surface charges, the apparent decrease in conductance in the dilute limit is due to electroneutrality breakdown. 

In the narrow nanopore (Fig.~\ref{fig_Compare}a) the behavior is more complicated. First, a small energy barrier of $\approx2\kbt$ is needed to obtain the correct trend. This small energy barrier is expected due to a  lower dielectric constant inside the nanopores, related to the confinement of water. More importantly, the fitted mobility, in this case, is significantly higher: $12$ times higher than bulk value for the lower two surface charges, and more than a $100$ times greater for the high surface charge (Fig.~\ref{fittedMobilities}). The large mobility in the high surface charge limit can be due to ion-ion correlations, where the positive charges push each other to move faster. It can also be related to enhanced water flow, due to an increased slip length \cite{holt2006fast,whitby2008enhanced}.

\subsection{Sub-nanometer channels}
We conclude by applying our model to sub-nanometer channels, which is the relevant limit for ion channels in nature. The most prominent example of a biological nanochannel is the Gramicidin-A channel. With a pore diameter of about $4$\AA, it is truly a one-dimensional system. It is often described using Michaelis-Menten type conductance: ions which travel through a channel that connects reservoirs A and B are transitioning between three possible states (``A", ``B" and ``in channel"). This framework is successful since it naturally ignores any charge neutrality constraints. It predicts a linear dependence in dilute solutions and a saturation at high concentrations, limited by the maximal occupancy of the pore. Despite its good agreement with experimental data, it can only describe systems with a handful of ions. Continuum models, on the other hand, which are much better suited to handle numerous ions, cannot predict the dilute limit linear behavior as long as electroneutrality is assumed. 

\begin{figure}[!]
  \includegraphics[width=\columnwidth ]{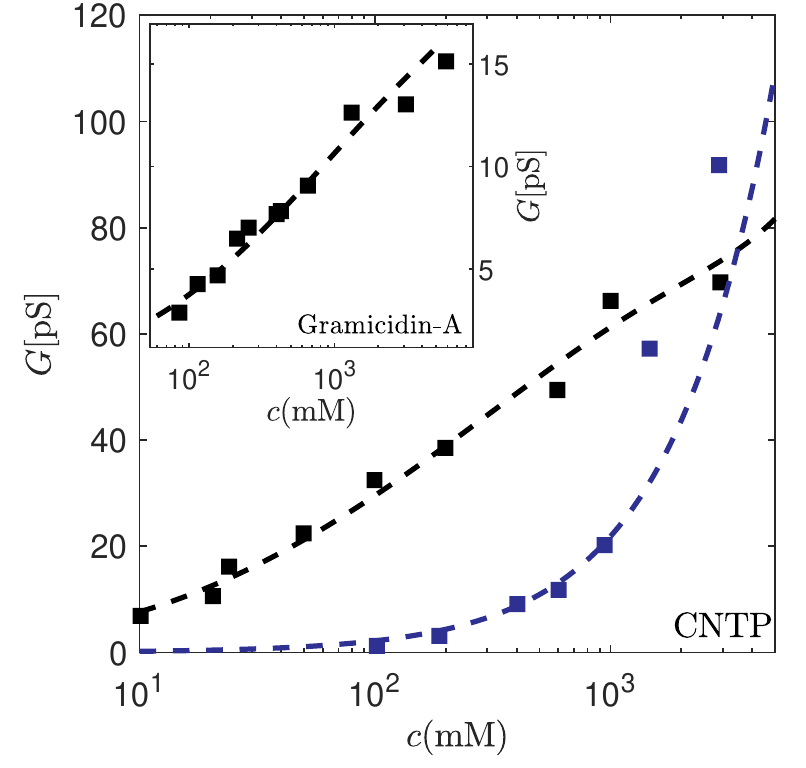}
  \caption{Conductance of sub-nanometer pores as a function of ionic concentration. Main figure: Conductance of a $3.5$\AA wide CNT porin as a function of KCl concentration (filled squares), adapted from \cite{Tunuguntla2017} and fitted according to our model (dashed lines). The bottom curve (blue) has pH 3.5, which corresponds to zero charge, while the upper (black) curve has pH 7.5, and is fitted to a surface charge of $6\mCpA$. Inset: conductance in Gramicidin-A channel as a function of NaCl concentration (filled squares, adapted from \cite{Finkelstein1981}) and its fit (dashed lines). The fitted surface charge is $13\mCpA$. }
  \label{gAfit}
\end{figure}

We consider two experimental datasets: a conductance measurement Gramicidin-A channel (taken from \cite{Finkelstein1981}), and more recent conductance measurement in a CNT porin experiment (taken from \cite{Tunuguntla2017}). We fit the data according to Eq.~\ref{eqnG} as before, with one important modification. As water molecules are excluded from these channels, the dielectric constant is now much smaller and was chosen to be $\varepsilon=5$. 


As shown in Fig.~\ref{gAfit}, our model is able to capture both neutral (linear conductance) and charged (Michalis-Menten conductance) sub-nanometer nanopore behavior. The plateau at high concentrations is not predicted, as our model fails in the concentrated regime. For higher concentrations, a more detailed picture of the (coupled) fluxes has to be accounted for and is beyond the scope of this paper. As shown by the lattice model (see Fig.~\ref{fig_1dPD}), ion-ion correlations can lead to a decrease in the total charge with increased concentration and eventually to an overall reduction in conductance.  

The predicted energy barriers for both experiments were similar ($5\kbt$). This energy is much smaller than the Born solvation energy in the vacuum, which might imply that dehydration is not complete,  and is compensated by the interactions of ions with the pore walls.   It is also smaller compared with energy barriers estimated by Michalis-Mentan type theories, which are of the order of $10\kbt$\cite{hille1978ionic}. We note, however, that in order to keep the model simple, we assumed a constant energy difference between the pore and its surrounding. Entrance effects were smeared throughout the system, so a smaller energy barrier is expected. A more careful derivation is required to accurately separate the pore mouth contribution to transport.

\begin{figure}[!]
  \includegraphics[width=\columnwidth ]{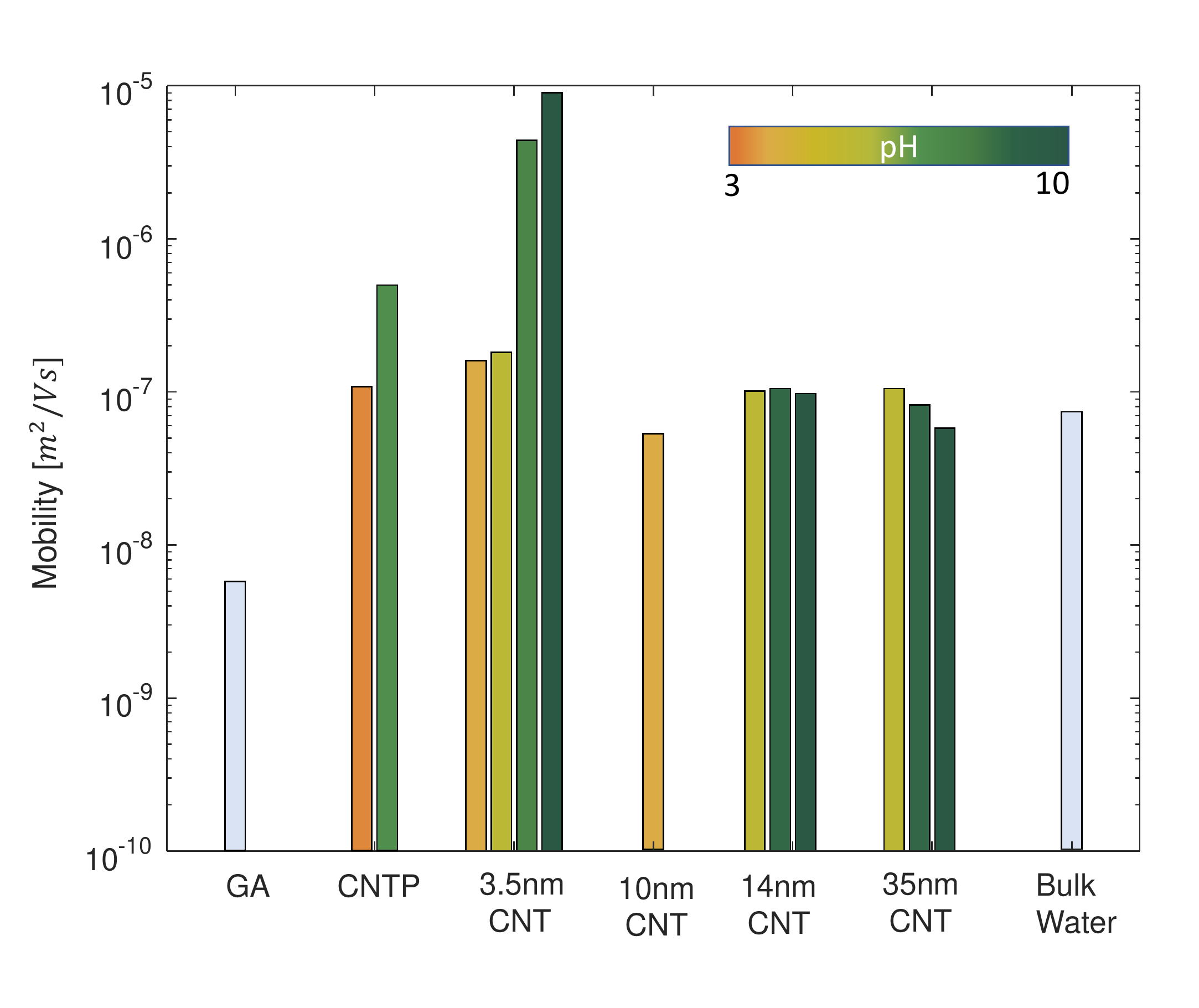}
  \caption{Fitted mobilities. The mobilities of $6$ datasets, Gramicidin-A, CNT Porin, and 4 different CNT experiments, were fitted according to Eq.~\ref{eqnG}. Experimental data and fits are shown if Figs.~7-8. Each pH was fitted separately, and within each experiment the bins are ordered from low pH to high. The bulk value for KCl mobility (last column) was taken from \cite{lide2004crc}}. 
  \label{fittedMobilities}
\end{figure}

Another interesting result is that the ionic mobility in GA is much lower compared to bulk water, while the CNT porins have higher mobility. The biological channel has fitted mobility of $5.7\times10^{-9}{\rm m}^2/{\rm V s}$ which is an order of magnitude less than the bulk ionic mobility (see Fig.~\ref{fittedMobilities}). The CNTP fitted mobility is closer to bulk KCl and equals $11\times10^{-9}{\rm m}^2/{\rm V s}$ and $50\times10^{-9}{\rm m}^2/{\rm V s}$ for the neutral and charged pores, respectively.   The reduced mobility compared to isolated ions in a solvent (water) can be understood from the strong attraction of ions and the opposite fixed wall charge and hence larger friction for relative motion. Our expression for conductance is derived based on a normal Nernst-Einstein relation with uncoupled fluxes of each ion. This is only valid for “pseudo binary” transport where each specie only interacts with an abundant solvent. Generally, the Stefan-Maxwell picture has a diffusion tensor with coupled fluxes between each pair of species \cite{donev2018fluctuating,balu2018role}, and can be extended to a ``dusty gas model," if wall molecules are treated as a fixed specie \cite{evans1961gaseous,fu2015multicomponent}.

\section{Conclusions}
We have analyzed ions in a $1$d electrolyte interacting with $3$d electrostatic interactions, which is the appropriate limit for single-digit nanopores. In $3$d systems the fluctuations around electroneutrality are limited only to the microscopic length-scale. The strong Coloumbic cost of large-scale deviations is much greater than the thermal energy. In contrast, when ions are forced to reside along a line, even macroscopically-long charged chains can spontaneously form. 

We first showed directly, using numerical simulations, that a net charge inside long and narrow nanopores is easily achieved when the radius of the pore is small compared to Debye length. We then developed an analytical mean-field model and and predicted a phase diagram for the accumulated charge inside the pore. We found that the pore behavior depends on two length-scales: the Debye screening length and the Gouy-Chapman length. If both length-scales are larger than the pore diameter, the ions do not know about the $3$d nature of the system, and electroneutrality is broken.  This behavior is best understood by examining how the competition between electrostatic forces and entropy determines the screening length. In three dimensions we recover the classical short-range Debye screening length, but in $1$d we find an exponentially large screening length. 

In our efforts to provide analytical results with a clear physical meaning, we neglected several important aspects of the problem. Most notably, our models fail to account for the polarization charge induced by a discontinuity of the dielectric constant. A large mismatch in the dielectric constant can alter ion-ion interactions inside the pore\cite{levin2006electrostatics,bordin2012ion}, and can lead, for example, to a $1$d Coulomb interactions in short nanopores\cite{zhang2005conductance}.  Molecular dynamics studies have shown that the interactions with images forces are especially relevant for selectivity in ion channels \cite{boda2006effect,boda2007steric,boda2007combined}. Selectivity is also sensitive to the size of the ions, and a proper theory of confined electrolytes must include finite size effects. 

Within the uniform embedded pore model, we derived approximated but accurate closed-form expressions for the expected charge and ionic concentration inside a charged pore, and the resulting ionic conductance. With two fitting parameters, the surface charge and ionic mobility, we were able to fit a wide range of conductance curves in narrow nanopores. We interpret the unusual scaling behavior observed as a consequence of the breakdown of electroneutrality. 

While transport measurements are an extremely useful tool for studying nanopores, inferring the ionic concentrations is a difficult task. Ion-ion, ion-water and water-pore interactions all play a role in the complicated transport phenomena. A more complete description that explicitly accounts for water flow is required to correctly predict the conductance. Moreover, the mobility of the ions may vary under confinement and composition\cite{esfandiar2017size,duan2010anomalous}, and the linear relation with concentration is only appropriate in infinite dilution.

Another important aspect of ionic transport, especially relevant to short nanopores, are entrance effects  and access resistance. The transition from a microchannel to a nano-channel adds additional resistance to the system, and the access resistance decreases with increasing concentration, which is an alternative explanation for the scaling observed in the conductance, as argued in recent papers\cite{green2014effect,green2016interplay}.  The entrance effects are of even greater importance if charges are added to the pore mouth, for example, to increase selectivity\cite{corry2011water,chan2013zwitterion,gravelle2013optimizing}. 

 With current available data, we cannot rule out alternative explanations to describe the scaling and shape of the conductance curves. However, as concentration decreases, our predictions deviate substantially from other models. For example, we predict a linear conductivity in the very dilute limit, and not a plateau. Experiments with a wider range of dilute concentrations are therefore needed to correctly identify the key mechanism.     

Added complexities are surely required to adequately describe the transport of ions through nanopores. Yet, as extremely long and narrow nanopores become technologically accessible, determining their net charge is a crucial first step. In our quest to understand the physics of single-digit nanopores, we highlight a simple but consequential observation on the nature of geometrical confinement: it breaks charge neutrality.

\begin{acknowledgments}
This research was supported as part of the Center for Enhanced Nanofluidic Transport
(CENT), an Energy Frontier Research Center funded by the U.S. Department of Energy,
Office of Science, Basic Energy Sciences under Award $\#$ DE-SC0019112 (continuum modeling), and by an Amar G. Bose Research Grant ($1$d lattice-gas model). JPD acknowledges support from the National
Science Foundation Graduate Research Fellowship under
Grant No. 1122374. 
\end{acknowledgments}

\bibliography{refs}

\appendix

\section{Exact solution of a charged cylinder in the Debye-H\"uckel regime} 
In section II.A we found an analytical solution for a weakly charged sphere in the DH regime. We show here the solution for a weakly charged cylinder, in the form of an infinite sum. We consider a cylinder of length $L$, radius $R$ filled with an electrolyte with permittivity $\varepsilon_{\rm in}$ and immersed in a dielectric medium dielectric constant $\varepsilon_{\rm out}$. The set of Debye-H\"uckel and Laplace equations, derived by linearizing Eq.~\ref{PBL_eqns}, are:
\begin{eqnarray}
\label{DHL_eqns}
\begin{cases}
    \nabla^2 \phi_{\rm in}(\rbf) =  \lambda_D^{-2} \phi_{\rm in}(\rbf) & r<R  \\
    \nabla^2 \phi_{\rm out} (\rbf) = 0 & r>R.
\end{cases}
\end{eqnarray}
The boundary conditions (Eq.~\ref{PBL_BC}) in a cylindrical geometry read:
\begin{eqnarray}
\label{PBL_BC_Cylinder}
\left[\phi_{\rm out}(R,z)-\phi_{\rm in}(R,z)\right] &=& 0  \nonumber\\ 
\left[\varepsilon_{\rm out}\partial_r \phi_{\rm out}(R,z)-\varepsilon_{\rm in}\partial_r \phi_{\rm in}(R,z)\right]  &=&  -q_s  (0<z<L) \nonumber\\
\left.\phi_{\rm out}(\rbf)\right|_{r\rightarrow \infty }&=&0 \nonumber\\
\phi_{\rm in}(R,z=0)=\phi_{\rm out}(R,z=0)&=&0\nonumber\\ \phi_{\rm in}(R,z=L)=\phi_{\rm out}(R,z=L)&=&0. 
\end{eqnarray}
Following \cite{Barcilon1992a}, we also imposed that the potential vanishes at the ends of the cylinder. Taking advantage of the azimuthal symmetry, we can write the inner and outer solutions as an infinite sum:
\begin{eqnarray}
    \phi_{\rm in}(r,z)&=&\sum_{n=1}^{\infty}A_n \sin(\omega_n z)I_0(\sqrt{\omega_n^2+\kappa_D^{2}}r)\nonumber\\
    \phi_{\rm out}(r,z)&=&\sum_{n=1}^\infty B_n \sin(\omega_n z)K_0(\omega_n r), 
\end{eqnarray}
where $I_0(x)$ and $K_0(x)$ are the $0^{\rm th}$ order modified Bessel function of first and second kind, respetively, and $\omega_n = \pi n/L$. Note that we wrote the solution in a manner that satisfies the last three boundary conditions. To find $A_n$ and $B_n$, we plug in the expansion to the first two boundary conditions, where the second boundary condition is expanded in a similar way: 
\begin{equation}
    \left[\varepsilon_{\rm out}\partial_r \phi_{\rm out}(R,z)-\varepsilon_{\rm in}\partial_r \phi_{\rm in}(R,z)\right] = - q_s\sum_{n \,{\rm odd}} \frac{4 \sin(\omega_n z)}{\omega_n L}.
\end{equation}
Clearly, for even $n$ we have $A_n=B_n=0$. For odd $n$, we obtain the following equations for the expansion coefficients:
\begin{eqnarray}
    A_n I_0(\sqrt{\omega_n^2+\kappa_D^{2}}R)&=&B_n K_0(\omega_n R),
    \nonumber\\
       \varepsilon_{\rm in}\sqrt{\omega_n^2+\kappa_D^{2}}  I_1(\sqrt{\omega_n^2+\kappa_D^{2}}R)  &=& -B_n \varepsilon_{\rm out} \omega_n  K_1(\omega_n R) + \frac{4 q_s}{\omega_n L}. \nonumber\\
\end{eqnarray}
Finally, we get:
\begin{widetext}
\begin{eqnarray}
\label{exactSol}
    \phi_{\rm in}(r,z)&=&\frac{2 q_s}{L}\sum_{n \, {\rm odd}}  \frac{1}{\omega_n}\frac{K_0(\omega_n R) \sin(\omega_n z)I_0(\sqrt{\omega_n^2+\kappa_D^{2}}r)}{\varepsilon_{\rm out}\sqrt{\omega_n^2+\kappa_D^{2}}K_0(\omega_n R)I_1(\sqrt{\omega_n^2+\kappa_D^{2}}R)+\varepsilon_{\rm out}\omega_n I_0(\sqrt{\omega_n^2+\kappa_D^{2}}R) K_1(\omega_n R)}.\nonumber\\
    \phi_{\rm out}(r,z)&=&\frac{2 q_s}{L}\sum_{n \, {\rm odd}}  \frac{1}{\omega_n}\frac{K_0(\omega_n r) \sin(\omega_n z)I_0(\sqrt{\omega_n^2+\kappa_D^{2}}R)}{\varepsilon_{\rm out}\sqrt{\omega_n^2+\kappa_D^{2}}K_0(\omega_n R)I_1(\sqrt{\omega_n^2+\kappa_D^{2}}R)+\varepsilon_{\rm out}\omega_n I_0(\sqrt{\omega_n^2+\kappa_D^{2}}R) K_1(\omega_n R)}.\nonumber\\
\end{eqnarray}
\end{widetext}

\section{Approximated boundary conditions for long and narrow nanopores} 

In the limit of small aspect ratios ($L\gg R$) we find an approximated Robin type boundary condition for the Poisson-Boltzmann equation, that obviates the need for outer fields solution. We do so by relating the electrostatic potential on the cylinder walls ($\phi_{\rm in}(R,z)=\phi_{\rm out}(R,z)$) to the outer electric field, $\varepsilon_{\rm out} \partial_r \phi_{\rm out}(R,z)$. This will allow us to express the jump in the electric field as a function of inner electric field and inner electrostatic potential, instead of solving the outer region as well. 

Since the outer region satisfies the Laplace equation, and as long as we have azimuthal symmetry, the electrostatic potential at the surface can be written as the following sum: 
\begin{equation}
    \phi_{\rm out}(R,z)=\sum_{n=1}^\infty B_n \sin\left(n\pi \frac{z}{L} \right)K_0\left( n\pi \frac{R}{L}\right), 
\end{equation}
Note that we are not limited to the linearized DH region as in in Appex.~A. In the limit $L\gg R$, the modified Bessel function is approximated to lowest order: 
\begin{eqnarray}
    \phi_{\rm out}(R,z) &=& -\sum_{n=1}^\infty B_n \sin\left(n\pi \frac{z}{L} \right) \left(\log\left(\frac{n \pi R}{2 L}\right)+\gamma_{\rm Euler}\right)
    \nonumber\\
    &+& O\left(\left(\frac{R}{L}\right)^2 \log\left(\frac{R}{L}\right)\right),
\end{eqnarray}
where $\gamma_{\rm Euler}\approx0.577$ is Euler's constant. We note that to lowest order in $\log(L/R)$ we find:
\begin{equation}
    \log\left(\frac{n \pi R}{2 L}\right) = \log\left(\frac{\pi R}{2 L}\right) + \log(n) \approx \log\left(\frac{\pi R}{2 L}\right). 
\end{equation}
which leads to the following approximation for the electrostatic potential: 
\begin{eqnarray}
    \phi_{\rm out}(R) &\approx& M_{L/R} \sum_{n=1}^{\infty} B_n \sin(\omega_n z), 
\end{eqnarray}
where $M_{L/R}$ is defined as:
\begin{equation}
    M_{L/R} = \log\left(\frac{2 L}{\pi R}\right)-\gamma.
\end{equation}
Our next step is to relate the derivative of the potential to the potential itself, which will allow us to mask the solution as a boundary condition. Taking the first derivative of the electrostatic potential we obtain:
\begin{equation}
        \partial_r\phi_{\rm out}(R,z) = -\sum_{n=1}^\infty B_n \omega_n K_1 (\omega_n R) \sin(\omega_n z). 
\end{equation}
Luckily, we find that to lowest order ($K_1(x)\approx 1/x$) it is indeed proportional to the electrostatic potential:
\begin{equation}
    \partial_r\phi_{\rm out}(R,z) \approx -\sum_{n=1}^\infty B_n  \frac{\sin(\omega_n z)}{R} = -\frac{\phi_{\rm out}(R,z)}{M_{L/R} R}. 
\end{equation}
Finally, assuming a surface charge on the pore walls ($q_s$), we get a new boundary condition for nanopores: 
\begin{equation}
\label{BC}
    \partial_r\phi_{\rm in}(R)=\frac{q_s}{\varepsilon_{\rm in}}- \frac{\varepsilon_{\rm out}}{\varepsilon_{\rm in}}\frac{\phi_{\rm in}(R)}{R M_{L/R}} .
\end{equation}

\end{document}